\documentclass[aps,prb,twocolumn,10pt,nofootinbib,superscriptaddress]{revtex4-2}
\usepackage{libertine}
\usepackage[T1]{fontenc}
\usepackage[libertine,upint]{newtxmath}
\usepackage[cal=stixtwofancy,frak=stixtwo,bb=stixtwo]{mathalfa}
\usepackage{graphicx,microtype,mathtools,bm,xcolor,caption,subcaption,floatrow}
\definecolor{linkscolor}{cmyk}{0.6, 0.3, 0, 0.9}
\usepackage[colorlinks=true,allcolors=linkscolor!60]{hyperref}
\usepackage{comment}
\usepackage{float}
\newcommand*{\ve}{\varepsilon}

\newcommand*{\vfst}{\varv^*_{\mathrm{F}}}
\newcommand*{\vfa}{\varv_{\mathrm{F},1}}
\newcommand*{\vfb}{\varv_{\mathrm{F},2}}

\newcommand*{\kfa}{k_{\mathrm{F},1}}
\newcommand*{\kfb}{k_{\mathrm{F},2}}
\newcommand*{\Nfa}{N_{\mathrm{F},1}}
\newcommand*{\Nfb}{N_{\mathrm{F},2}}

\newcommand*{\bk}{{\bm{k}}}
\newcommand*{\bp}{{\bm{p}}}

\newcommand*{\bq}{{\bm{q}}}

\newcommand*{\nn}{\nonumber}
\newcommand*{\beq}{\begin{equation}}
\newcommand*{\eeq}{\end{equation}}
\newcommand*{\bea}{\begin{eqnarray}}
\newcommand*{\eea}{\end{eqnarray}}
\newcommand*{\bse}{\begin{subequations}}
\newcommand*{\ese}{\end{subequations}}
\newcommand*{\bwt}{\begin{widetext}}
\newcommand*{\ewt}{\end{widetext}}
\newcommand*\im{{\mathrm{Im}}}

\newcommand*{\bv}{{\bm{v}}}

\DeclareMathOperator{\I}{\mathrm{Im}}
\DeclareMathOperator{\R}{\mathrm{Re}}

\newcommand*{\bsu}{\begin{subequations}}
\newcommand*{\esu}{\end{subequations}}

\newcommand*{\ofl}{\omega_{\text{FL}}}

\newcommand*{\Eq}{Eq.~\eqref}

\newcommand*{\bi}{\begin{itemize}}
\newcommand*{\ei}{\end{itemize}}

\newcommand*{\mB}{q_{\text{B}}}
\newcommand*{\qfa}{q_{\text{F},1}}
\newcommand*{\qfb}{q_{\text{F},2}}
\newcommand*{\cnd}{c^{\phantom{\dagger}}}

\DeclarePairedDelimiter\abs{\lvert}{\rvert}

\NewDocumentCommand{\grad}{e{_^}}{%
    \mathop{}\!
    \bm{\nabla}
    \IfValueT{#1}{_{\!#1}}
    \IfValueT{#2}{^{#2}}
}

\begin{document}

\title{Quantum criticality and optical conductivity in a two-valley system}

\author{Yasha Gindikin}
\email{gindikin@protonmail.ch}
\affiliation{W.I.\ Fine Theoretical Physics Institute and School of Physics and Astronomy, University of Minnesota, Minneapolis, Minnesota 55455, USA}
\author{Songci Li}
\affiliation{Department of Physics and Tianjin Key Laboratory of Low-Dimensional Materials Physics and Preparation Technology, Tianjin University, Tianjin 300354, China}
\author{Alex Levchenko}
\affiliation{Department of Physics, University of Wisconsin-Madison, Madison, Wisconsin 53706, USA}
\author{Alex Kamenev}
\author{Andrey V.\ Chubukov}
\affiliation{W.I.\ Fine Theoretical Physics Institute and School of Physics and Astronomy, University of Minnesota, Minneapolis, Minnesota 55455, USA}
\author{Dmitrii L.\ Maslov}
\affiliation{Department of Physics, University of Florida, Gainesville, Florida 32611-8440, USA}

\begin{abstract}
We demonstrate that the optical conductivity of a Fermi liquid (FL) in the absence of umklapp scattering is dramatically affected by the topology of the Fermi surface (FS). Specifically, electron-electron (\emph{ee}) scattering leads to rapid current relaxation in systems with multiple, or multiply connected, FSs, provided that the valleys have different effective masses. This effect results from intervalley drag. We microscopically derive the optical conductivity of a two-valley system, both within the FL regime and near a quantum critical point (QCP) of the Ising-nematic type. In the FL regime, intervalley drag restores the Gurzhi-like scaling of the  conductivity, $\R\sigma(\omega) \sim \omega^0$. This dependence contrasts sharply with the previously identified sub-leading contribution to the conductivity of  a two-dimensional  FL with a single convex FS, where $\R\sigma(\omega) \sim \omega^2 \ln |\omega|$. The vanishing of the leading term in the optical conductivity is a signature of geometric constraints on \emph{ee} scattering channels, which are lifted for a multiply connected FS. A large differential response, $d\R\sigma/d \mu$ with $\mu$ being the chemical potential, is predicted at the Lifshitz transition from a single-valley to a multi-valley FS, which should be observable within the experimentally accessible frequency range. Near a QCP, intervalley drag leads to a $|\omega|^{-2/3}$ scaling of $\R\sigma(\omega)$ in 2D, thus providing a specific current-relaxing process for this long-standing conjecture.
\end{abstract}
\maketitle

\section{Introduction}

In recent years there has been a revival of interest to optical studies of various materials with a non-trivial band structure, including graphene 
in different incarnations~\cite{Peres:2010,Gmitra:2015}, transition metal dichalcogenides (TMD)~\cite{manzeli:2017,heinz:2018}, and Weyl and Dirac semimetals \cite{Hosur:2013,Armitage:2018b}. The latest advancements in optical spectroscopy, encompassing a broad frequency spectrum, have provided a unique probe of intra-band electron dynamics, demonstrating how electron-electron (\textit{ee}) interactions influence current relaxation and manifest themselves in distinct patterns of frequency scaling of the conductivity~\cite{ivchenko2005optical,basov:2005,basov:2011,klingshirn2012,maslov:2017b,Armitage:2018,Tanner:book}. To a large degree, these studies are driven by an enduring interest in the formation of novel quantum states.

The dissipative part of the optical conductivity of a non-Galilean-invariant Fermi liquid (FL) at frequencies $\omega \tau \gg 1$, where $\tau \propto T^{-2}$ is the quasiparticle transport time, is believed to be described by the Gurzhi formula~\cite{gurzhi:1959},
\begin{equation}
	\label{eq:Gurzhi}
	\R \sigma(\omega,T) =\sigma_0 \left(1 + \frac{4\pi^2T^2}{\omega^2}\right)\,,
\end{equation}
with $\sigma_0$ being the material-dependent constant.
However, recent studies have demonstrated that, in the absence of umklapp scattering, the validity of the Gurzhi scaling is critically dependent on the Fermi surface (FS) geometry and topology~\cite{rosch:2005,rosch:2006,maslov:2017b,Sharma:2021,Guo:2022,Li:2023,Guo:2023,Gindikin:2024}. Specifically, an isotropic FS both in two-dimensional (2D) and three-dimensional (3D) systems, and a convex 2D FS gives rise to a dramatic suppression of the conductivity due to a slow current relaxation.\footnote{In 2D, a slow current relaxation is the manifestation of a more general effect: a long lifetime of odd harmonics of the electron distribution function~\cite{gurzhi:1980,gurzhi:1982,gurzhi:1987,gurzhi:1995,levitov:2019}.} This unfolds as follows.

Regardless of the method employed — be it equations of motion for the current operator~\cite{rosch:2005,rosch:2006,Sharma:2021,Li:2023}, Fermi Golden Rule (FGR)~\cite{mishchenko:2004,Sharma:2024}, or diagrammatic expansion of the Kubo formula~\cite{maslov:2017b,Gindikin:2024} — the optical conductivity is ultimately proportional to $(\Delta \bm{v})^2$, averaged over the FS with the scattering probability, where
\begin{equation}
	\Delta \bm{v} \equiv \bm{v}_{\bm{p}} + \bm{v}_{\bm{p}'} - \bm{v}_{\bm{k}} - \bm{v}_{\bm{k'}}
\end{equation}
is a change of the total group velocity (proportional to the total current) due to a collision of two electrons with initial momenta $\bm{k}$ and $\bm{k}'$, and final momenta $\bm{p}$ and $\bm{p}'$. In a Galilean-invariant system, $\bm{v}=\bm{k}/m$, and $\Delta\bm{v}$ is equal to zero identically by  momentum conservation. Same happens for an isotropic FS both in 2D and 3D, except for in this case $\Delta\bm{v}$ vanishes only on the FS rather than identically~\cite{rosch:2005,rosch:2006,Sharma:2021}. For a convex FS in 2D, the only allowed scattering channels are the Cooper channel, in which $\bk=-\bk'$ and $\bp=-\bp'$, and the swap channel, in which $\bk'=\bp$ and $\bp'=\bk$, and $\Delta\bv=0$ for both these channels~\cite{gurzhi:1980,gurzhi:1982,gurzhi:1987,gurzhi:1995,maslov:2011,pal:2012b}. As long as $\Delta\bv=0$, the conductivity vanishes to leading order for all the cases described above and, to obtain a finite result, one needs to expand $\Delta\bv$ either around the FS (for an isotropic FS) or near the Cooper and swap solutions (for a convex FS). This leads to a strong suppression of the conductivity compared to the Gurzhi formula: $\R\sigma(\omega,T)\propto \max\{\omega^2\ln|\omega|^{-1},T^4\ln T^{-1}/\omega^2\}$ in 2D and   $\R\sigma(\omega,T)\propto \max\{\omega^2,T^4/\omega^2\}$ in 3D.

The Gurzhi result for the conductivity is restored under specific conditions. In 3D, this occurs when anisotropic terms in the electron dispersion beyond the quadratic level are taken into account. In 2D, the Gurzhi result is valid for a concave FS, which allows for scattering channels beyond the Cooper and swap ones with $\Delta\bv\neq 0$~\cite{maslov:2011,pal:2012b,Li:2023,Gindikin:2024}.

In this work, we consider another mechanism of fast, Gurzhi-like current relaxation: \emph{ee} scattering in systems with multiple, or multiply connected, FSs with different effective masses. At the phenomenological level, the case of a multivalley conductor with parabolic dispersions in each valley can be analyzed via the classical equations of motion, in which the intervalley interaction is accounted for by the drag force, proportional to the difference of the drift velocities in different valleys. The drag force renders the \textit{ee} contribution to the conductivity finite, both in the optical case and in the \emph{dc} case in the presence of disorder~\cite{levinson:book,maslov:2017b}. The \emph{dc} case was further analysed in Refs.~\cite{maslov:2011} and~\cite{pal:2012b} via the semiclassical Boltzmann equation.  The goal of the present paper is to derive the optical conductivity of a two-valley system microscopically, both in the FL regime and in the situation when the \emph{ee} interaction drives the system to the quantum critical point (QCP) of the Ising-nematic type~\cite{fradkin:2010}.

We start with a simpler problem of a two-valley FL with Coulomb interaction within and between two nonequivalent valleys. Our model spans a wide range of 2D systems and materials, from systems with non-parabolic spectrum and valley imbalance, typical for TMDs, to systems with annular FS, including biased bilayer and rhombohedral trilayer graphene. We calculate the optical conductivity via the FGR and show that it exhibits the Gurzhi scaling of Eq.~\eqref{eq:Gurzhi}. The interest to such systems lies in the ability to manipulate the scattering channels by electrical means, which promises large differential responses at the threshold of the channel opening, thus crafting a unique experimental signature within the experimentally accessible frequency range.

We proceed with a more intricate example of a two-valley system, in which one of the valleys is tuned to an Ising-nematic QCP, while the other valley plays the role of a momentum sink for the first one. We demonstrate that the intervalley conductivity scales with frequency as $\abs{\omega}^{-2/3}$ at the QCP in 2D, which is the same as conjectured a long time ago in Ref.~\cite{kim:1994} in the context of fermions interacting with a $U(1)$ gauge field.

\section{Two-valley Fermi liquid away from criticality}
\label{sec:away}
\subsection{The model}
We start with an effective Hamiltonian of a two-valley electron system\footnote{Extending our approach to accommodate an arbitrary number of valleys is conceptually straightforward.}
\begin{equation}
    \label{eq:ham}
        H = \sum_{\bm{k},s} (\varepsilon_{\bm{k}, s}-\mu) c_{\bm{k}, s}^{\dagger} c_{\bm{k}, s}^{\vphantom{\dagger}} + \frac12 \sum_{\bm{q}s s'} u_{s s'}(\bm{q}) \rho_{\bm{q},s} \rho_{-\bm{q},s'}\,.
\end{equation}
Here, $c_{\bm{k}, s}$ and $\rho_{\bm{q}, s}$ represent the electron annihilation operator and the density operator, respectively, $\varepsilon_{\bm{k}, s}$ is the electron dispersion, $\mu$ is the chemical potential, and indices $s,s'=1,2$ label distinct valleys. We ignore the electron spin if the valleys do not result from spin splitting; if they do, $s$ and $s'$ label the spin projection. The Hamiltonian captures two types of \emph{ee} interactions involving small momentum transfers: intravalley interaction, denoted by $u_{ss}(\bm{q})$, and intervalley drag, denoted by $u_{ss'}(\bm{q})$, but neglects the intervalley swaps of electrons. The last assumption is justified if the valley centers are located far away from each other or, for the case of concentric valleys, if their Fermi momenta are widely different. For simplicity, we will consider the case of isotropic but otherwise arbitrary $\ve_{\bk,s}$; adding anisotropy of the FS does not change the scaling form of $\R\sigma(\omega,T)$. We focus on the 2D case first and discuss the 3D case at the end of this section.

The distinction between $u_{11}$, $u_{22}$ and $u_{12}$ becomes significant near the QCP, a topic explored in detail in the Section~\ref{sec:QCP}. Here, away from criticality, we simplify the model by assuming uniformity of the interactions: $u_{ss}(\bm{q}) = u_{ss'}(\bm{q}) = \mathcal{U}_{\bm{q}}$. In the static limit, we consider the 2D screened Coulomb interaction in the Thomas-Fermi approximation,
\begin{equation}
    \mathcal{U}(\bm{q}) = \frac{2 \pi e^2}{\abs{\bm{q}}+ \kappa},\; \kappa = 2\pi e^2(\Nfa+\Nfb)\,,
\end{equation}
with $N_{\mathrm{F},s}$ being the density of states at the Fermi energy in valley $s$. We assume that the interaction is sufficiently long-ranged, such that the condition to neglect intervalley swaps is satisfied.

The gradient part of the current operator is given by
\begin{equation}
	\label{eq:normal_current}
	\bm{J} = \sum_{\bm{k}, s} \bm{v}_{\bm{k}, s} c_{\bm{k}, s}^{\dagger} c_{\bm{k}, s}^{\vphantom{\dagger}},\quad \bm{v}_{\bm{k}, s} = \grad_{\bm{k}} \varepsilon_{\bm{k}, s} = \hat{k} \, v_{\bm{k}, s}\,,
\end{equation}
where $\hat{k} \equiv \bm{k}/k$.

The real part of the conductivity, $\sigma (\omega)$, can be derived via the FGR (see Refs.~\cite{mishchenko:2004,Sharma:2024} and Appendix~\ref{app:FGR}) up to the two-loop order to give
\begin{widetext}
    \begin{align}
    \label{eq:FGR}
        \R \sigma(\omega,T)
          = {}& \frac{\pi e^2}{2 \omega D}\left(1-e^{-\frac{\omega}{T}}\right) \sum_{\stackrel{\bm{k} \bm{k'} \bm{q}}{s s'}}
    		\int\! d\Omega\, \abs{\bm{\mathfrak{M}}_{ss'}(\bm{k},\bm{k}',\bm{q})}^2 n\left(\varepsilon_{\bm{k}+\tfrac{\bm{q}}{2},s}-\Omega\right)
              n\left(\varepsilon_{\bm{k}'-\tfrac{\bm{q}}{2},s'}-\omega+\Omega\right)
              \left(1-n\left(\varepsilon_{\bm{k}+\tfrac{\bm{q}}{2},s}\right)\right)
              \left(1-n\left(\varepsilon_{\bm{k}'-\tfrac{\bm{q}}{2},s'}\right)\right)\notag \\
    		&{} \times \delta\left(\Omega+\varepsilon_{\bm{k}-\tfrac{\bm{q}}{2},s}-\varepsilon_{\bm{k}+\tfrac{\bm{q}}{2},s}\right) \delta\left(\omega -\Omega +\varepsilon_{\bm{k}'+\tfrac{\bm{q}}{2},s'}-\varepsilon_{\bm{k}'-\tfrac{\bm{q}}{2},s'}\right)\,,
    \end{align}
\end{widetext}
where $\bm{\mathfrak{M}}_{ss'}(\bm{k},\bm{k}',\bm{q})$ is the matrix element of the Coulomb interaction and $n(\varepsilon_{\bm{k}, s})={\left(e^{\frac{\varepsilon_{\bm{k}s}-\mu}{T}}+1\right)}^{-1}$ is the Fermi function, and $D$ is the spatial dimensionality. Note that the conductivity represents the sum of intravalley, $\sigma_{ss}$, and intervalley, $\sigma_{s\ne s'}$, contributions, of which we are mostly interested in the latter.

For our case of a long-range interaction, the exchange part of the matrix element can be neglected, and then
\begin{equation}
\label{eq:matrix_el}
	\bm{\mathfrak{M}}_{ss'}(\bm{k},\bm{k}',\bm{q}) = \frac{\mathcal{U}(\bm{q}) \Delta \bm{v}_{ss'}}{\omega}\,,
\end{equation}
where
\begin{equation}
	\label{eq:velocity_change}
	\Delta \bm{v}_{s s'} \equiv \bm{v}_{\bm{k}+\tfrac{\bm{q}}{2},s} + \bm{v}_{\bm{k}'-\tfrac{\bm{q}}{2},s'} - \bm{v}_{\bm{k}'+\tfrac{\bm{q}}{2},s'} - \bm{v}_{\bm{k}-\tfrac{\bm{q}}{2},s}
\end{equation}
is the velocity change due to a two-electron scattering process in a two-valley system. In contrast to the single-valley case, $\Delta \bm{v}_{s s'}\neq 0$ even for the parabolic dispersions $\ve_{\bk,s}=k^2/2m_s$, as long as $m_1\neq m_2$.

\subsection{Fermi surface geometry}

The velocity change in Eq.~\eqref{eq:velocity_change} can be expanded, to the leading order in the momentum transfer $q$, into the sum of two qualitatively different terms, $\Delta \bm{v}_{ss'} = \Delta \mathfrak{u}_{ss'} + \Delta \mathfrak{w}_{ss'}$, where
\begin{equation}
	\label{eq:velocity_change_1}
	\Delta \mathfrak{u}_{ss'} = \bm{q}
 \left( \frac{1}{\bar m_s(k)}-\frac{1}{\bar m_{s'}(k')}\right)\,,
\end{equation}
and
\begin{equation}
\label{eq:velocity_change_2}
 \begin{split}
	\Delta \mathfrak{w}_{ss'} = {}&
	\left(\frac{1}{m^*_s(k)}-\frac{1}{\bar m_s(k)}\right)
	\hat{k} (\bm{q} \cdot \hat{k})\\
	&- \left(\frac{1}{m^*_{s'}(k')}-\frac{1}{\bar m_{s'}(k')}\right) \hat{k}' (\bm{q} \cdot \hat{k}')\,,
 \end{split}
\end{equation}
with $1/\bar m_s(k)=(1/k)\partial \ve_{\bk,s}/\partial k$ and $1/m^*_s(k)=\partial^2\ve_{\bk,s}/\partial k^2$ being the (inverse) density-of-states and band masses, respectively.

We call two valleys ``equivalent'',  if they have equal density-of-states masses on the corresponding FSs, i.e.,
$\bar{m}_1\equiv \bar m(\kfa)=\bar m_2\equiv \bar m(\kfb)$. For equivalent valleys, $\Delta \mathfrak{u}_{ss'}$ gives zero when projecting $k$ and $k'$ onto the
corresponding FSs. In this case, the main contribution to the conductivity comes from $\Delta \mathfrak{w}_{ss'}$ but, similarly to the single-valley case, this contribution is suppressed as compared to the Gurzhi result of Eq.~\eqref{eq:Gurzhi}.\footnote{Note that $\Delta \mathfrak{w}_{ss'}=0$ for a parabolic dispersion.}

For nonequivalent valleys the situation is quite the opposite — the dominant contribution to the conductivity is due to $\Delta \mathfrak{u}_{ss'}$, and this dramatically affects the scaling form of the conductivity.

\subsection{Optical conductivity}
\label{sec:FGL}

The dissipative conductivity, to the leading order of the interaction strength, is determined by all possible \emph{ee} scattering processes accompanied by the creation of two electron-hole pairs.

Energy conservation in a process of \emph{ee} scattering is ensured by two $\delta$-functions in Eq.~\eqref{eq:FGR}:
\begin{equation}
	\label{eq:en_conserv}
	\begin{split}
		&\varepsilon_{\bm{k}+\tfrac{\bm{q}}{2},s}-\varepsilon_{\bm{k}-\tfrac{\bm{q}}{2},s}=\Omega\,,\\
		&\varepsilon_{\bm{k}'-\tfrac{\bm{q}}{2},s'}-\varepsilon_{\bm{k}'+\tfrac{\bm{q}}{2},s'}=\omega-\Omega\,.
	\end{split}
\end{equation}
To leading order in $\max\{\omega,T\}$, one can set the right-hand sides of the last two equations to zero. Then, at fixed $\bq$, the allowed electron momenta $\bm{k}$ satisfy the equation $\varepsilon_{\bm{k}-\tfrac{\bm{q}}{2}} = \varepsilon_{\bm{k}+\tfrac{\bm{q}}{2}}$. Geometrically, the solutions correspond to the intersection points between two FSs, shifted by the momentum transfer $\bm{q}$.

\begin{figure}[htb]
	\includegraphics[width=.5\linewidth]{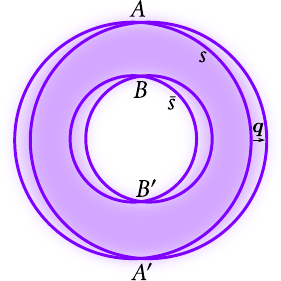}
	 \caption{The intersection points between two Fermi surfaces, shifted by the momentum transfer $\bq$. \label{fig:fs}}
\end{figure}

In what follows we assume that the electron dispersion is even in $\bk$, i.e., $\ve_{\bk,s}=\ve_{-\bk,s}$. The time-reversal symmetry guarantees this to be the case if spin-orbit interaction can be neglected, regardless of whether the inversion symmetry is present or not~\cite{Bir1974}. If so, the intersection points arise in pairs at ${\bm k}$ and $-{\bm k}$, like at points $A$ and $A'$ in Fig.~\ref{fig:fs}.

The dominant contribution to the conductivity at the smallest energy transfer $\Omega$ (which is of the same order as the photon frequency $\omega$)
comes from \emph{ee} scattering in a close vicinity of the intersection points. A \textit{channel of electron-electron scattering} is defined by any pair of the intersection points. The contributions from distinct channels are being summed up in the expression for the conductivity.

For a single circular (and, more general, convex) FS in 2D, there are at most two intersections between the shifted FSs. In this case, the only scattering channels for electrons are the swap channel with $\bm{k} \approx \bm{k}'$ (channels $\{A,A\}$ and $\{A',A'\}$ in Fig.~\ref{fig:fs}) and Cooper channel with $\bm{k} \approx -\bm{k}'$ (channels $\{A,A'\}$ and $\{A',A\}$ \textit{ibid}.)

The situation changes qualitatively for electrons residing in nonequivalent valleys. The new scattering channels arise for two electron situated near the intersection points of distinct FSs: $\{A,B\}$, $\{A,B'\}$, etc. When two electrons are scattered close to these points, their momenta align in the same or opposing directions, $\hat{k} \approx \hat{k}'$ or $\hat{k} \approx -\hat{k}'$, but the absolute values of their momenta are no longer equal, $\abs{\bm{k}} \ne \abs{\bm{k}'}$, in a distinction from the swap and Cooper channels.

The intervalley ($s'\ne s$) contribution to the conductivity of Eq.~\eqref{eq:FGR} is determined by the factor $\Delta \bm{v}_{ss'}$, evaluated
for $\bm{k}$ and $\bm{k}'$ on the corresponding FSs. In this case, the leading contribution to $\Delta \bm{v}$ comes from $\Delta \mathfrak{u}_{ss'}=\bm{q}\delta$ of Eq.~\eqref{eq:velocity_change_1}, with the mass mismatch
\begin{equation}
	\delta = \frac{v_{Fs}}{k_{Fs}} - \frac{v_{Fs'}}{k_{F s'}} = \frac{1}{\bar{m}_s} - \frac{1}{\bar{m}_{s'}}\,.
\end{equation}
For nonequivalent valleys, $\delta \ne 0$.

It is crucially important that, in contrast to $\delta \mathfrak{w}_{ss'}$ of Eq.~\eqref{eq:velocity_change_2}, $\Delta \mathfrak{u}_{s s'}$ is not small when electron momenta are close to their respective intersection points. This smallness, inherent in Eq.~\eqref{eq:velocity_change_2}, reflects the small-angle nature of the \textit{ee} scattering, and is enforced by the dot product of $\bm{q}$ and $\hat{k}$,
\begin{equation}
	(\bm{q} \cdot \hat{k}) \approx v_{ks}^{-1}\left(\varepsilon_{\bm{k}+\tfrac{\bm{q}}{2},s}-\varepsilon_{\bm{k}-\tfrac{\bm{q}}{2},s}\right)\,,
\end{equation}
which, according to Eq.~\eqref{eq:en_conserv}, gives an extra power of frequency to conductivity (see Eq.~\eqref{eq:Maslov} below).

Hence, the intervalley conductivity is given by
\begin{equation}
	\begin{split}
		\R \sigma_{s\neq s'} (\omega) = {}&\frac{e^2}{4 \pi \omega^3}(1-e^{-\frac{\omega}{T}}) \sum_{\bm{q}}\int d\Omega\, \mathcal{U}_{\bm{q}}^2 {(\Delta \mathfrak{u}_{s s'})}^2 \\
		&\times \im \mathfrak{P}_s(\Omega,\bm{q}) \im \mathfrak{P}_{s'}(\omega-\Omega,\bm{q})\,,
	\end{split}
\end{equation}
with
\begin{equation}
	\mathfrak{P}_s(\Omega,\bm{q}) = \sum_{\bm{k}} \frac{n(\varepsilon_{\bm{k}+\tfrac{\bm{q}}{2},s}-\Omega)(1-n(\varepsilon_{\bm{k}+\tfrac{\bm{q}}{2},s}))}{\Omega + \varepsilon_{\bm{k}-\tfrac{\bm{q}}{2},s} -\varepsilon_{\bm{k}+\tfrac{\bm{q}}{2},s}+i0}\,.
\end{equation}
Note that $\mathfrak{P}_s$ is distinct from the electron polarization $\Pi_s$, which arises from the interplay of ``absorption'' and ``emission'' processes, represented as $n(1-n')-n'(1-n)$. In contrast, $\mathfrak{P}_s$ only involves the ``absorption'' component. Consequently, its imaginary part is not an odd function of the frequency.

In the limit of $v_{\mathrm{F},s} q \gg \abs{\Omega}$, one gets
\begin{equation}
	\im \mathfrak{P}_s(\Omega,\bm{q}) = \frac{N_{\mathrm{F,s}}}{e^{-\frac{\Omega}{T}}-1}\frac{\Omega}{v_{\mathrm{F},s} q}\,.
\end{equation}

A subsequent integration over $\Omega$ yields
\begin{equation}
	\label{eq:freq_int}
		\int_{-\infty}^{\infty} \frac{\Omega(\omega-\Omega)\,d\Omega}{\left(e^{-\frac{\Omega}{T}}-1\right)\left(e^{-\frac{\omega-\Omega}{T}}-1\right)} = \frac{\omega^3}{6\left(1-e^{-\frac{\omega}{T}}\right)}\left(1+\frac{4\pi^2 T^2}{\omega^2}\right)\,.
\end{equation}

In 2D, the integral over momentum transfer is regularized by imposing the ultra-violet cutoff $\lambda=\min\{k_{Fs},k_{F s'}\}$ to give
\begin{equation}
	\int q \mathcal{U}^2_{\bm{q}}\, dq = {(2\pi e^2)}^2\left[\ln \left(1+\frac{\lambda}{\kappa}\right) - \frac{\lambda}{\lambda+\kappa}\right]\,,
\end{equation}
where the expression in brackets, $\mathcal{C}$, interpolates between
$\mathcal{C} =\ln{(\lambda/\kappa)}$ for $\lambda \gg \kappa$, and $\mathcal{C} =\lambda^2/(2\kappa^2)$ for $\lambda \ll \kappa$.

This way, we obtain the intervalley conductivity in 2D,
\begin{equation}
	\label{eq:optconduct}
	\sigma_ {s\neq s'}(\omega) = \frac{\mathcal{C} \delta^2}{12} e^2  \frac{e^2 N_{\mathrm{F,s}}}{v_{\mathrm{F},s}} \frac{e^2 N_{\mathrm{F,s'}}}{v_{\mathrm{F},s'}} \left(1+\frac{4\pi^2 T^2}{\omega^2}\right) \,.
\end{equation}
Remarkably, the signature of the FL state — the Gurzhi scaling of Eq.~\eqref{eq:Gurzhi} — is reinstated due to the drag between nonequivalent valleys.

The 3D case can be considered along the same lines. Although a generic 3D FS allows for infinitely many channels of current relaxation, the intravalley contribution to the conductivity still vanishes to leading order for the special case of an isotropic FS. Indeed, the group velocity in this case is given by $\bv_{\bk,s}=\bk/\bar m_s(k)$. Projecting all momenta in Eq.~\eqref{eq:velocity_change} onto the FS, we obtain $\Delta\bv_{ss}\vert_{k=k_{\mathrm{F},s}}=0$. A finite result is obtained by expanding $\Delta\bv_{ss}$ near the FS, and the resultant conductivity is suppressed by a factor of $\max\{\omega^2,T^4/\omega^2\}$ \cite{Sharma:2021,Goyal:2023}. The intervalley contribution differs from the 2D result \eqref{eq:optconduct} only by a numerical coefficient due to a different form of the screened Coulomb interaction.

\subsection{Applications}

Consider a two-valley Dirac metal with the dispersion of $\varepsilon_{\bm{k}s} = v\abs{\bm{k}}$ and valley imbalance — be it interaction-driven or invoked by the Zeeman splitting — characterized by $\zeta= \Delta \varepsilon_F/\mu=|\ve_{\mathrm{F},1}-\ve_{\mathrm{F},2}|/\mu$. In this case the total intervalley conductivity $\sigma=\sigma_{12}+\sigma_{21}$ becomes
\begin{equation}
	\label{eq:mc}
	\R \sigma(\omega) = e^2\frac{\alpha^2 \abs{\ln{\alpha}}}{24 \pi^2} \zeta^2 \left(1+\frac{4\pi^2 T^2}{\omega^2}\right) \,,
\end{equation}
with the effective fine structure constant $\alpha=e^2/v$.

The intravalley contribution~\cite{Sharma:2021,Goyal:2023}
\begin{equation}
	\label{eq:Maslov}
	\R \sigma_{ss}(\omega) = \frac{e^2}{960 \pi^2} \frac{\omega^2}{\ve_{\mathrm{F},s}
 ^2} \left(1+\frac{4\pi^2 T^2}{\omega^2}\right)
	\left(3+\frac{8\pi^2 T^2}{\omega^2}\right) \ln \frac{\alpha k_{\mathrm{F},s}}{\Lambda}\,,
\end{equation}
with infrared cutoff $\Lambda = \max \left\{\omega/v,T/v\right\}$, is suppressed by a factor of $\max\{\omega^2 \ln \omega,T^4\ln T/\omega^2\}$, as compared to Eq.~\eqref{eq:mc}.

The next example is given by a biased bilayer graphene, which features a moat-band dispersion of the form~\cite{McCann_2013}
\begin{equation}
\label{eq:moat_band}
	\varepsilon_{\bm{k}} = \frac{V}{2} - V \frac{v^2}{t_{\perp}^2}k^2 + \frac{v^4}{t_{\perp}^2 V}k^4\,,
\end{equation}
where $t_{\perp}$ the interlayer hopping parameter and $V$ the voltage drop across the layers. Provided that a single band is occupied, a multiply-connected annular FS, depicted in Fig.~\ref{fig:fs}, arises if $V>2\mu$.

Here, the mass mismatch between the inner and outer components of the FS equals
\begin{equation}
	\delta = \frac{4 v^4}{t_{\perp}^2 V}(k_{Fs}^2 + k_{Fs'}^2) \approx \frac{4 v^2 V}{t_{\perp}^2}\,,
\end{equation}
and $\mathcal{C} \approx (V-2\mu) t_{\perp}^2/4\kappa^2 v^2 V$, for $\mu$ close to $2V$. This gives rise to the conductivity exhibiting a threshold behavior, described by the following expression:
\begin{equation}
	\label{eq:bbg}
    \R \sigma(\omega) = \mathcal{D}\, \theta(V-2\mu) (V-2\mu) \left(1+\frac{4\pi^2 T^2}{\omega^2}\right)\,,
\end{equation}
where $\mathcal{D}$ represents a combination of the system parameters. This threshold behavior marks the Lifshitz transition, at the voltage $V$ equal to $2\mu$.~\footnote{While the model predicts a sharp transition at this threshold, it is important to note that at finite temperature, this transition is not perfectly abrupt. Specifically, we assume that the temperature $T$ is much smaller than the Fermi energies $T\ll \min\{\varepsilon_{F,1},\varepsilon_{F,2}\}$, yet in the vicinity of the transition it becomes large enough to introduce a smearing around the threshold on the scale of $T$.}   

The valence band of biased rhombohedral trilayer graphene also features a
moat-band dispersion. While the low-energy Hamiltonian fully accounting for the moat-band structure is yet to be derived from the many-band model~\cite{PhysRevB.107.104502}, for current purposes it is sufficient to approximate it by the dispersion similar to Eq.~\eqref{eq:moat_band}, again neglecting the trigonal warping effect. Then we obtain
\begin{equation}
	\label{eq:trg}
	\R \sigma(\omega) = \tilde{\mathcal{D}}\, \theta(2\mu-V) (2\mu-V) \left(1+\frac{4\pi^2 T^2}{\omega^2}\right)\,.
\end{equation}

In modern double-gate experimental setups, it is possible to independently adjust the voltage across layers without changing the electron concentration. This allows for the electrical control of conductivity, according to Eqs.~(\ref{eq:bbg}--\ref{eq:trg}), at fixed electron density. Opening a new scattering channel within the small FS by adjusting the chemical potential to cross the peak of the ``Mexican hat'' gives rise to the jump in the differential conductivity, which could be directly observed through modulation spectroscopy~\cite{voigtlander2015}.

\section{Quantum-critical two-valley system}
\label{sec:QCP}
\subsection{The Model}
\subsubsection{Charge currents}
In this section, we consider a two-valley system near an Ising-nematic quantum critical point.
The interaction part of the  Hamiltonian is the same as in \Eq{eq:ham}, except for the form-factors that 
project
the interaction onto a channel with a given angular momentum. Accordingly, $u_{ss'}(\bq) \to F_s(\bk) F_{s'}(\bp)u_{ss'}(\bq)$. Due to these form-factors, the electric field couples not only to the free but also to the interaction part of the Hamiltonian. Correspondingly, the current contains the regular part, given by \Eq{eq:normal_current}, and the ``anomalous'' part~\cite{Li:2023,Gindikin:2024}:
\bea
\bm{J}_{\text{an}}&=&
\sum_{\bk, \bp, \bq, s,s'}\left[
\left(\grad_{\bk} +\grad_\bp\right)F_s (\bk) F_{s'} (\bp) \right]u_{ss'}(\bq)\nn\\
&&\times c^\dagger_{\bk+\frac{\bq}{2},s}c^\dagger_{\bp-\frac{\bq}{2},s'}\cnd_{\bp+\frac{\bq}{2},s'}\cnd_{\bk-\frac{\bq}{2},s}.\label{currents}
\eea

The time derivatives of the current operators in Eqs.~\eqref{eq:normal_current} and~\eqref{currents} show directly which processes contribute to current relaxation. For the two-valley case, we obtain
\bwt
\bse
\bea
\partial_t\bm{J}
=-i[H,\bm{J}]
=-i[H_{\text{int}},\bm{J}
]=-\frac{i
}{2}
\sum_{\bk,\bp,\bq, s,s'}&&(\bv_{\bk+\bq/2,s}+\bv_{\bp-\bq/2,s'}-\bv_{\bk-\bq/2,s}-\bv_{\bp+\bq/2,s'})
F_s(\bk)F_{s'}(\bp)u_{ss'}(\bq)\nn\\&&\times c^\dagger_{\bk+\bq/2,s}
c^\dagger_{\bp-\bq/2,s'} \cnd_{\bp+\bq/2,s'} \cnd_{\bk-\bq/2,s}
\label{K1}
\eea
and
\bea
 \partial_t\bm{J}_{\text{an}}=-i[H,\bm{J}_{\text{an}}]=-i[H_0,\bm{J}_{\text{an}}]=
 i
 \sum_{\bk, \bp, \bq, s, s'}&&\left(\ve_{\bk+\bq/2,s}+\ve_{\bp-\bq/2,s'}-\ve_{\bk-\bq/2,s}-\ve_{\bp+\bq/2,s'}\right)(\grad_{\bk}+\grad_{\bp}) F_s(\bk) F_{s'}(\bp)u_{ss'}(\bq)\nn\\
 &&\times c^\dagger_{\bk+\bq/2,s}  c^\dagger_{\bp-\bq/2,s'} \cnd_{\bp+\bq/2,s'} \cnd_{\bk-\bq/2,s}.\label{K2}
\eea
\ese
\ewt

We see that  $\partial_t\bm{J}_{\text{an}}$ in Eq.~\eqref{K2} contains the difference between the energies of two electrons before and after a collision, which is equal to the photon energy, $\omega$. Therefore, the corresponding contribution to the conductivity is suppressed by a factor of $\omega^2$ \cite{Li:2023,Gindikin:2024}. On the contrary, $\partial_t\bm{J}$ in Eq.~\eqref{K1} contains the difference of the velocities of the initial and final states. This difference is non-zero for non-degenerate valleys, even if the electron spectrum in each of the valleys is parabolic, and the corresponding contribution to the conductivity is not suppressed. For this reason, we will neglect the anomalous part of the current. For the normal part of the current, the form-factors do not bring any qualitative changes, and we will ignore them in what follows. Thus, the interaction part of the Hamiltonian is reduced back to \Eq{eq:ham}.

A FL near a QCP is strongly interacting, i.e., its $Z$-factor is much smaller than one (or, equivalently, the renormalized mass is much larger than the bare one). To keep track of the Z-factor renormalization, we employ the diagrammatic treatment of the Kubo formula in this section, as opposed to the FGR scheme employed in Sec.~\ref{sec:away}.

\subsubsection{Quantum criticality in a two-valley system}
\label{sec:QC2VS}
As in the previous part of the paper, we assume that valleys are located sufficiently far from each other, such that the exchange of fermions between the valleys can be neglected. We adopt a model of Hubbard-like interaction ($u_{ss'}=\mathrm{const})$ and, for simplicity, assume that fermions in valley 2 do not interact with each other ($u_{22}=0$). Then the only bare interactions in the model are the intravalley interaction in valley 1 $(u_{11})$ and intervalley drag ($u_{12}$). As before, we also assume that the electron spectrum in each valley is isotropic but not necessarily parabolic, but consider both the 2D and 3D cases on the same footing.

Within the Random Phase Approximation (RPA), the matrix of dressed interaction is given by\footnote{RPA is controllable if the number of fermionic flavors in each valley, $N$, is large. We assume this to be the case but, for brevity, do not display $N$ in the formulas.}
\bea
\hat U
=
\hat u\left(\hat{\mathbb{1}}-
\hat\Pi
\hat u
\right)^{-1},\label{RPA}
\eea
where

\bea
\Pi_{ss'}(\bq,\Omega_m)&=&T\sum_{\nu_m} \int \frac{d^D k}{(2\pi)^2} 
G_s(\bk+\bq/2,\nu_m+\Omega_m/2)\nn\\
&&\times G_{s'}(\bk-\bq/2,\nu_m-\Omega_m/2)
\eea
and $G_s(\bk,\nu_m)$ is the Green's function of valley $s$.

The solution of Eq.~\eqref{RPA} for our case of $u_{22}=0$ is given by
 \bse
 \bea
  U_{11}  (\bq,\Omega_m)&=&
  \frac{u_{11}+\Pi_{22}(\bq,\Omega_m)}{R}
  u^2_{12}
  \nn\\
 &=& \frac{u_{11}}{R_1(\bq,\Omega_m)}
  +\frac{\Pi_{22}(\bq,\Omega_m)u^2_{12}}{R_1(\bq,\Omega_m)R(\bq,\Omega_m)},\label{U11g}\\
  U_{12}  (\bq,\Omega_m)&=& U_{21}  (\bq,\Omega_m)=\frac{u_{12}}{R(\bq,\Omega_m)},\label{U12g}\\
  U_{22}  (\bq,\Omega_m)&=&\frac{\Pi_{11}(\bq,\Omega_m)u_{12}^2}{R(\bq,\Omega_m)},\label{U22g}
 \eea
 \ese
 where
 \bse
\bea
R_1(\bq,\Omega_m)&=&1-u_{11}\Pi_{11}(\bq,\Omega_m),\label{R1}\\
R(\bq,\Omega_m)&=&R_1(\bq,\Omega_m)-\Pi_{11}(\bq,\Omega_m)\Pi_{22}(\bq,\Omega_m) u_{12}^2.\nn\\\label{R12}
\eea
\ese
The QCP in a two-valley system occurs when $R(\bq\to 0,\Omega_m=0)=0$, which corresponds to the condition
\bea
1+\lambda_{11}-\lambda_{12}=0,
\eea
where
\bse
\bea
\lambda_{11}&=&
\Nfa u_{11}\label{l11}\\
\mathrm{and}\;\lambda_{12}&=& \Nfa\Nfb u^2_{12}
\label{l12}
\eea
\ese
are the dimensionless couplings of the intra- and intervalley interaction, respectively, and $N_{\mathrm{F},s}$ is the density of states in valley $s$ at the corresponding Fermi energy.  For future convenience, we singled out the first term in Eq.~\eqref{U11g}, which is the  intravalley interaction in the absence of coupling between the valleys.

In what follows, we will need the asymptotic forms of the polarization bubbles for
\bea
|\Omega_m|\ll \varv_{\text{F},s}q\ll k_{\mathrm{F},s},\label{Oq}
\eea which are given by
\bea
\Pi_{ss}(\bq,\Omega_m)=-N_{F,s}\left(1-\frac{q^2}{q^2_{\mathrm{F},s}}-
C_D\frac{|\Omega_m|}{\varv_{\text{F},s} q}\right),\label{Pi11}
\eea
where $C_2=1$, $C_3=\pi/2$, and $q_{\text{F},s}\sim k_{\text{F},s}$.\footnote{A $q^2$ term in the free-fermion polarization bubble is present as long as the band dispersion is neither parabolic nor linear. In addition, the interaction between low-energy fermions generate extra $q^2$ terms~\cite{maslov:2017}. It is assumed that the $q^2$ term in Eq.~\eqref{Pi11} includes both the free-fermion and interaction-induced  parts.}

Now we focus on the immediate vicinity of the QCP, where $R(\bq,\Omega)$ in Eq.~\eqref{R1} is finite but small. According to Eqs.~\eqref{R1} and \eqref{R12}, the factor $1/R_1$ in Eq.~\eqref{U11g} can be then replaced by $1/\Pi_{11}(\bq,\Omega_m)\Pi_{22}u_{12}^2$, which in non-singular. Therefore, the first term in this equation can be neglected. Keeping only the most singular terms in the  rest of the interactions in Eq.~\eqref{U11g}-\eqref{U22g}, we cast them into a familiar $Z=3$ critical form \cite{hertz:1976}:
\bse
\bea
U_{11}(\bq,\Omega_m)&=&-\frac{1}{\Nfa R(\bq,\Omega_m)}\label{U11c}\\
U_{12}(\bq,\Omega_m)&=& U_{21}(\bq,\Omega_m)= \frac{u_{12}}{R(\bq,\Omega_m)}\label{U12c}\\
U_{22}(\bq,\Omega_m)&=& -\frac{\Nfa u_{12}^2}{R(\bq,\Omega_m)}\label{U22c}
\eea
\ese
with
\bea
R(\bq,\Omega_m)=\frac{1}{q^{*2}}\left(q^2+\mB^2+\gamma\frac{|\Omega_m|}{q}\right).\label{Rc}
\eea
Here, the parameters of the critical interaction are
\bea
\frac{1}{q^{*2}}&=&\frac{1}{\qfa^2}+\frac{\lambda_{12}}{\qfb^2},\;\frac{1}{\vfst}=\frac{1}{\vfa}+\frac{\lambda_{12}}{\vfb}\nn\\
\gamma&=& C_D\frac{q^{*2}}{\vfst},\;\mB^2=(1+\lambda_{11}-\lambda_{12})q^{*2},\label{param}
\eea
where $\mB$ is the mass (the inverse correlation length) of order parameter fluctuations. At the QCP, $\mB=0$. As long as $\mB>0$, the system is in the FL regime for $\omega\ll \ofl$ and in the non-Fermi-liquid (NFL) regime for  $\omega\gg\ofl$,  where
\bea
\ofl\equiv \vfst\mB^3/q^{*2}.
\eea

In the FL regime, the fermionic self-energy of the $s^{\text{th}}$ valley behaves as $\Sigma_{s}(\nu_m) \propto i\nu_m/\mB+i\mB^{-4}\nu
_m|\nu_m|\ln (\ofl/|\nu_m|)$ in 2D~\cite{Pimenov:2022}, and as $\Sigma_{s}(\nu_m)  \propto i\nu_m\ln (\ofl/|\nu_m|)+i\mB^{-3}\nu_m|\nu_m|$ in 3D.

A crossover to the NFL regime can be  achieved by employing the space-time scaling of the $Z=3$ critical theory~\cite{chubukov:2017}. This allows us to replace $\mB\to|\nu_m|^{1/3}$, which yields familiar results: $\Sigma_s(\nu_m)\propto i\mathrm{sgn}\nu_m|\nu_m|^{2/3}$ in 2D and $\Sigma_s(\nu_m)\propto i\nu_m\ln|\nu_m|$ in 3D. The justification for this step is as follows.

For a single-valley case, the conductivity, calculated via the fully dressed current-current correlation function, scales as $\sigma'(\omega) \propto \omega_{\mathrm{FL}}^{-2/3} F(\omega / \omega_{\mathrm{FL}})$. This result was derived in Ref.~\cite{chubukov:2017} under assumption that $\omega_{\mathrm{FL}}$ is the only energy scale near a nematic QCP, which is also the assumption we adopt in the current paper. The scaling function $F(x)$ is determined such that $\omega_{\mathrm{FL}}$ drops out from the result in the quantum-critical regime, where $\omega \gg \omega_{\mathrm{FL}}$. This leads to $F(x) \propto x^{-2/3}$ as $x \to \infty$. This behavior implies that in the vicinity of the QCP, the bosonic mass can be effectively replaced by $\omega^{1/3}$, with results in agreement with Refs.~\cite{kim:1994,eberlein:2016}.

We conjecture that this same scaling behavior applies to the two-valley case as well. However, this conjecture awaits confirmation through a detailed calculation of the fully dressed current-current correlation function. In the next section, we present a calculation of the optical conductivity in the FL regime near a QCP.

\subsection{Optical conductivity}
\subsubsection{Intravalley contribution}
\label{sec:intra}

\begin{figure}[htb]
\includegraphics[width=1.0\linewidth]{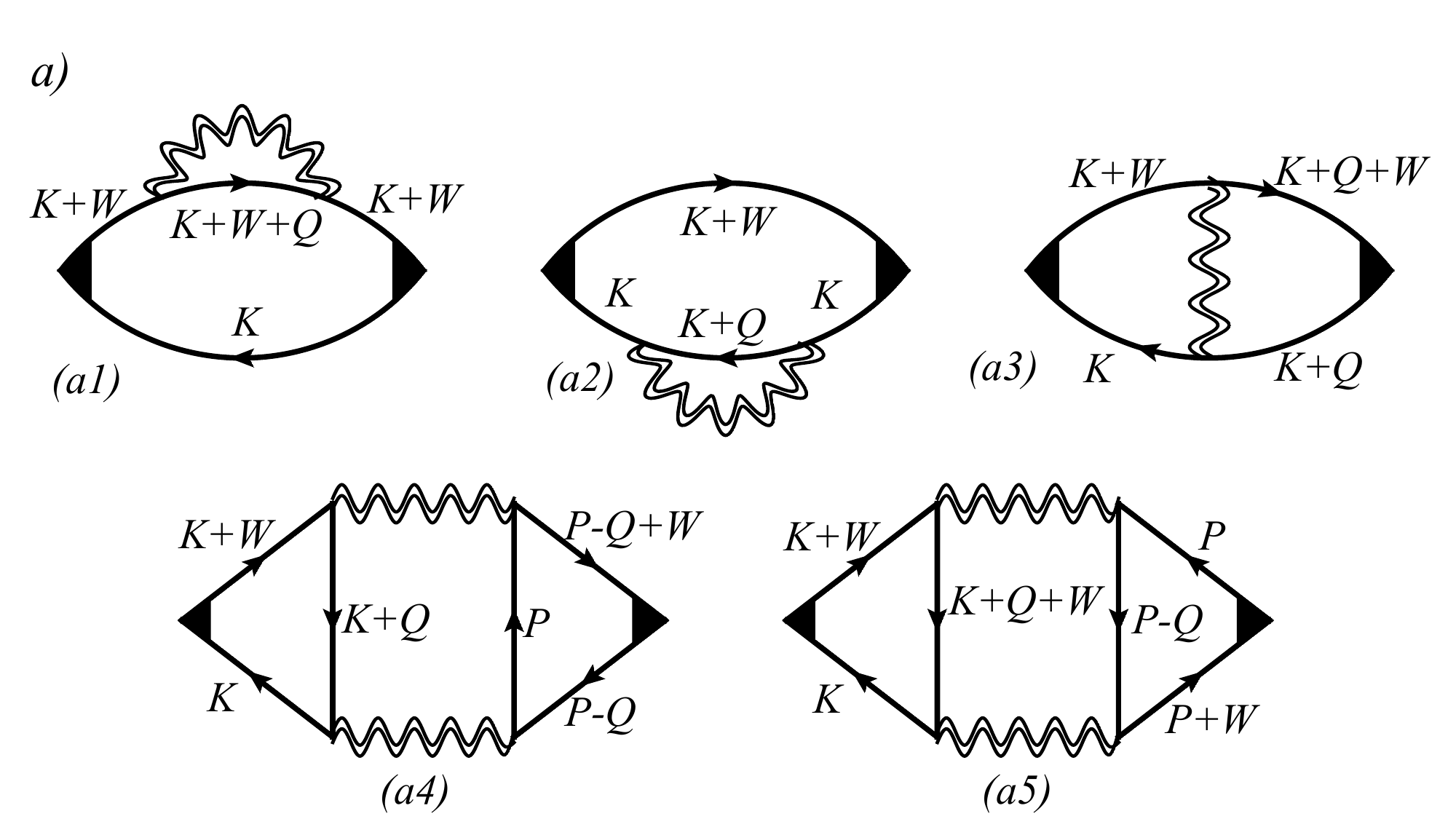}
\includegraphics[width=1.0\linewidth]{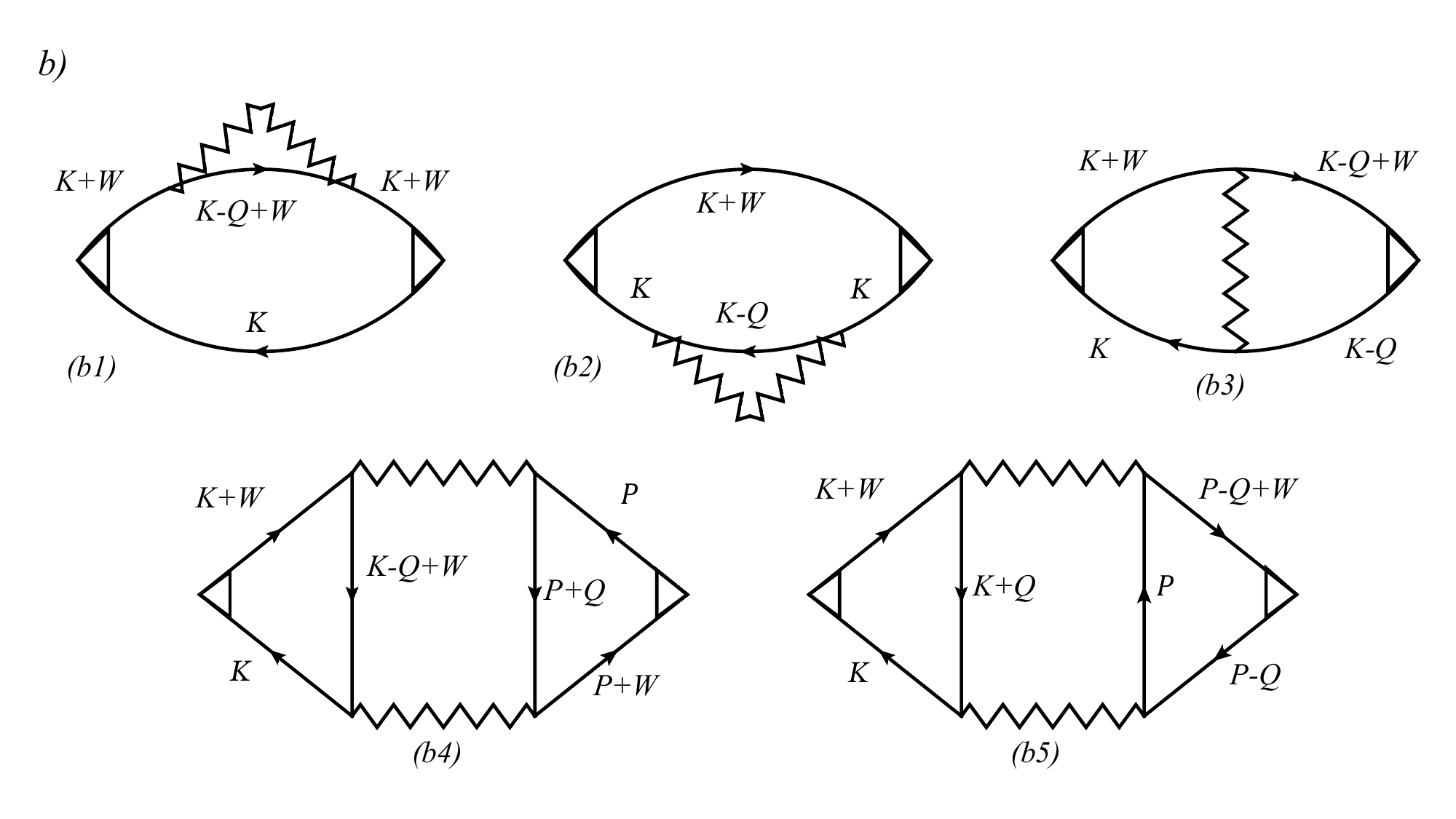}
 \includegraphics[width=1.0\linewidth]{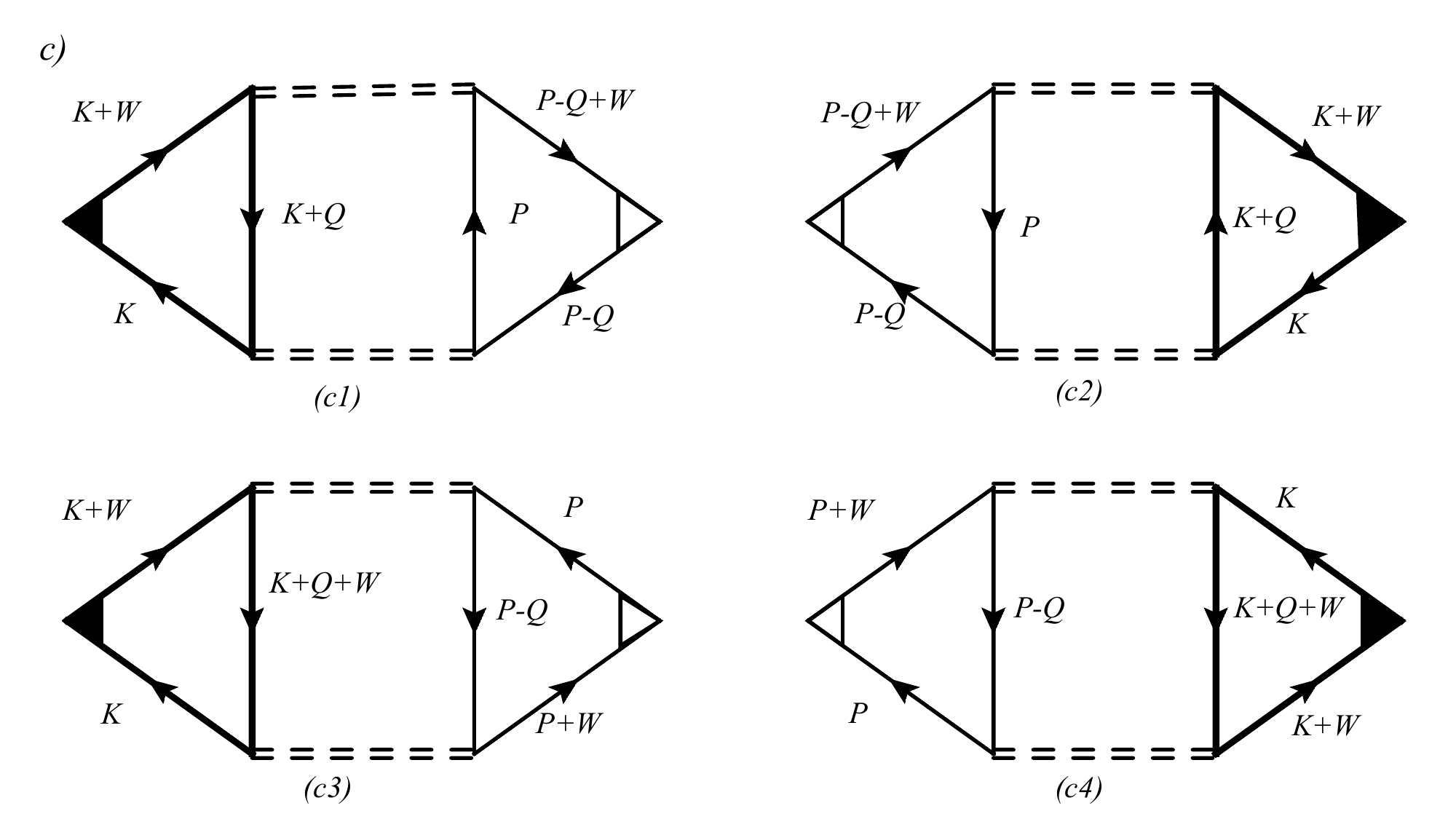}
		\caption{Diagrams for the optical conductivity of a two-valley quantum critical system.   a) Contribution from the intravalley interaction in valley 1. Thick solid line: Green's function of valley 1;  double wavy line: the $U_{11}$ component of the interaction, Eq.~\eqref{U11g}; solid triangle: renormalized current vertex. The capital letters label the $D+1$ momenta: $P=(\bp,\nu_m)$, $K=(\bk,\nu_m')$, $Q=(\bq,\Omega_m)$, $W=({\bf 0},\omega_m)$. b) Contribution from the induced intravalley interaction in valley 2.   Thin solid line: Green's function of valley 2; zig-zag line: $U_{22}$ component of the interaction, Eq.~\eqref{U22g}; open triangle: renormalized current vertex. c) Intervalley (drag) contribution. Double dashed line: the $U_{12}$ component of the interaction, Eq.~\eqref{U12g}.
		\label{fig:sigma11}}
\end{figure}

The leading-order diagrams for the contribution to the conductivity from the interaction between valley-1 and valley-2 fermions are shown in Figs.~\ref{fig:sigma11} \emph{a} and \emph{b}. The analysis of the single-valley case follows along the same lines as in Refs.~\cite{maslov:2017b,Sharma:2021,Li:2023,Gindikin:2024}, with additional details provided in Appendix~\ref{app:abc}.

The real part of the optical conductivity is non-zero only if the effective interaction is dynamic. For the self-energy (SE) and Maki-Thompson (MT) diagrams (\emph{a1-a3} and \emph{b1-b3} in Figs.~\ref{fig:sigma11}), it implies that one can subtract off the static part of the interaction, $U_{ss}^{\text{st}}(\bq)=U_{ss}(\bq,0)$, such that the double wavy and zigzag lines in these diagrams are replaced by the corresponding dynamic interactions,
\bea
U_{ss}^{\text{dyn}}(\bq,\Omega_m)=U_{ss}(\bq,\Omega_m)-U_{ss}(\bq,0),
\label{Udyn}\eea
expanded to linear order in $|\Omega_m|$.
In the FL regime, the Green's functions in all diagrams can be approximated by their quasiparticle forms\footnote{There is no double counting in adding self-energy corrections to the renormalized Green's functions in diagrams \emph{a1} and \emph{a2}, and \emph{b1} and \emph{b2}. Indeed, Eq.~\eqref{Gf} accounts only for the real part of the self-energy, while the double wavy line in these diagrams produces, after analytic continuation, the imaginary part of the self-energy.}
\bea
G_s(\bk,\nu_m)=\frac{1}{\frac{i\nu_m}{Z_s}-\ve_{\bk,s}},\,s=1,2,\label{Gf}
\eea
where $Z_s$ is the renormalization factor of valley $s$, and where we have taken into account that the theory becomes local near QCP, i.e., the self-energy depends primarily on the frequency but not on the momentum~\cite{Chubukov:2005self}.

Filled and blank triangles in Fig.~\ref{fig:sigma11} represent renormalized current vertices, $\bm{\Lambda}_{\bk,s}$.  For an isotropic spectrum and forward-type scattering, such a vertex can be written as the product of the bare group velocity and the charge vertex,
while the latter is related to the $Z$-factor via the Ward identity~\cite{chubukov:2017}:
\begin{eqnarray}
	\bm{\Lambda}_{\bk,s}=\bm{v}_{\bk,s}\Gamma_s=\frac{\bm{v}_{\bk,s}}{Z_s}.\label{Ward}
\end{eqnarray}

With Eqs.~\eqref{Gf} and~\eqref{Ward} taken into account,  the sum of the SE and MT diagrams is reduced to (cf. Appendix~\ref{app:abc})
\bwt
\bea
\sigma^{\text{SE+MT}}_{
ss}(\omega_m)= -\frac{e^2}{2D\omega^3_m}\sum_{K,Q}
(\bv_{\bk+\bq,s}-\bv_{\bk,s})^2
U_{ss}^{\text{dyn}}(Q)
\left[2G_s(K)G_s(K+Q)-G_s(K)G_s(K+Q+W)-G_s(K+Q)G_s(K+W)\right],\nn\\
 \label{sigma22a-c_main}
\eea
\ewt
where $K=(\bk,i\nu_m)$, $Q=(\bq,i\Omega_m)$, $W=({\bf 0},i\omega_m)$, $\sum_K\equiv {(2\pi)}^{D+1}\int d\nu_m \int d^D k$, and $G_s(K)\equiv G_s(\bk,\nu_m)$. Here and thereafter, we set $\omega_m>0$ without loss of generality.

In contrast to diagrams \emph{a1-a3} and \emph{b1-b3}, the interaction lines
in the Aslamazov-Larkin (AL) diagrams (\emph{a4,a5} and \emph{b4,b5} in Fig.~\eqref{fig:sigma11}) can be replaced by their static limit,
\bea
U_{ss}^{\text{st}}(\bq)=U_{ss}(\bq,\Omega_m=0),
\eea
while dynamics comes from the Green's functions forming the triangles.
The two AL diagrams can be combined into  the following expression (cf. Appendix~\ref{app:de})
\bwt
\begin{align}
\sigma^{\text{AL}}_{
ss}(\omega_m) = {} &\frac{e^2}{2D\omega^3_m}\sum_{K,Q}\sum_P
(\bv_{\bk+\bq,s}-\bv_{\bk,s})\cdot(\bv_{\bp,s}-\bv_{\bp-\bq,s})
\left[U^\text{st}_{ss}(\bq)\right]^2 G_s(P)G_s(P-Q)   \nn \\
	& \times \left[2G_s(K)G_s(K+Q)-G_s(K)G_s(K+Q+W)-G_s(K+Q)G_s(K+W)\right].
 \label{sigma22de_symm_MT}
\end{align}
\ewt

Combining Eqs.~\eqref{sigma22a-c_main} and \eqref{sigma22de_symm_MT}, and taking into account that typical momenta and energy transfers in the FL regime are in the range \eqref{Oq},
we obtain for the total intravalley contribution (see Appendix \ref{sec:comp})
\bwt
\begin{equation}
\sigma
_{
\mathrm{intra}
}(\omega_m) =  \sum_s\left(\sigma_{ss}^{\mathrm{SE+MT}}+ \sigma_{ss}^{\mathrm{AL}}\right)=\sum_s \frac{e^2C_DN_{\mathrm{F},s}\mathcal{O}_{D}}{2D\omega^3_m
\varv_{\mathrm{F},s}
\bar m_s^2}\int\frac{dqq^D}{(2\pi)^D}
\int\frac{d\Omega_m}{2\pi}
	\left(|\Omega_m+\omega_m|+|\Omega_m-\omega_m|-2|\Omega_m|\right)
U^{\mathrm{eff}}_{ss}(\bq,\Omega_m),
 \label{sigma22abc_4}
\end{equation}
\ewt
where
\begin{equation}
U^{\mathrm{eff}}_{ss}(\bq,\Omega_m)=U^{\text{dyn}}_{ss}(\bq,\Omega_m)-\left[U_{ss}^{\text{st}}(\bq)\right]^2\Pi_{ss}^{\text{dyn}}(\bq,\Omega_m)
\end{equation}
is the effective intravalley interaction,
\begin{equation}
	\Pi_{ss}^{\text{dyn}}(\bq,\Omega_m)=C_D N_{\mathrm{F},s} \frac{|\Omega_m|}{\varv_{\mathrm{F},s}q}
\end{equation}
is the dynamic part of the polarization bubble, and $\bar m_s=k_{\text{F},s}/\varv_{\text{F},s}$ is the density-of-states mass evaluated on the FS\@.
In the absence of the intervalley interaction, the intravalley interaction is reduced to the first term in Eq.~\eqref{U11g}. In this case, $U^{\mathrm{eff}}_{ss}(\bq,\Omega_m)$ vanishes and, to get a finite conductivity, one needs to expand the electron velocities near the FS. As a result, the conductivity is suppressed by a factor of $\omega_m^2\ln|\omega_m|$ in 2D and $\omega_m^2$ in 3D \cite{Sharma:2021,Guo:2022,Li:2023,Guo:2023,Gindikin:2024}. For the case of coupled valleys, however, $U^{\mathrm{eff}}_{ss}(\bq,\Omega_m)$ is finite because the dynamic part of the critical interactions in Eqs.~\eqref{U11c}-\eqref{U22c} comes from both valleys, while  $\Pi_{ss}^{\text{dyn}}(\bq,\Omega_m)$  comes only from one of the valleys. Explicitly, we obtain
\bwt
\bse
\bea
U^{\mathrm{eff}}_{11}(\bq,\Omega_m)&=&\frac{C_D}{N_{\mathrm{F},1}}\frac{q^{*4}}{(q^2+\mB^2)^2}\left(\frac{1}{\vfst}-\frac{1}{\varv_{\mathrm{F},1}}\right)\frac{|\Omega_m|}{q}=\frac{C_D\lambda_{12}}{N_{\mathrm{F},1}}\frac{q^{*4}}{(q^2+\mB^2)^2}\frac{|\Omega_m|}{\vfb q},\\
U^{\mathrm{eff}}_{22}(\bq,\Omega_m)&=&\frac{C_D\lambda_{12}}{N_{\mathrm{F},2}}\frac{q^{*4}}{(q^2+\mB^2)^2}\left(\frac{1}{\vfst}-
\frac
{1}{\varv_{\mathrm{F},2}}\right)\frac{|\Omega_m|}{q}=\frac{C_D\lambda_{12}}{N_{\mathrm{F},2}}\frac{q^{*4}}{(q^2+\mB^2)^2}\frac{|\Omega_m|}{\vfa q}.
\eea
\ese
\ewt
Note that the two intravalley interactions are related by a permutation of the valley indices $1\leftrightarrow 2$.

Performing straightforward integrations over $q$ and $\Omega_m$, and analytically continuing to real frequencies, we obtain the intravalley part of the conductivity as
\bea
\R\sigma_{\text{intra}}(\omega)&=&
A_De^2\lambda_{12}\frac{q^{*4}}{\vfa\vfb \mB^{4-D}}\left(\frac{1}{\bar m_1^2}+\frac{1}{\bar m_2^2}\right),\label{sigma11_final}
\eea
where $A_2=1/48\pi^2$ and $A_3=1/576$.

\subsubsection{Intervalley contribution}
\label{sec:intravalley}

The intervalley contribution to the conductivity is depicted graphically in Fig.~\ref{fig:sigma11}\emph{c}. Note that the  SE and MT  diagrams  are absent for the intervalley case, because they involve swapping of fermions between the valleys, which is not allowed in our model. Therefore, we have to consider only the AL diagrams. As with other AL diagrams, the full intervalley interaction (the double-dashed line) is replaced by its static value [cf. Eq.~\eqref{U12c}],
$U_{12}^{\text{st}}(\bq)=u_{12}q^{*2}/
(q^2+\mB^2)$.
As shown in Appendix~\ref{app:de}, the sum of diagrams in panel \emph{c} of Fig.~\ref{fig:sigma11} can be written as
\bwt
\bea
\sigma_{\text{inter}}(\omega_m)&=&\frac{e^2}{D\omega^3_m}\sum_{K,Q}\sum'_P
(\bv_{\bk+\bq,1}-\bv_{\bk,1})\cdot(\bv_{\bp,2}-\bv_{\bp-\bq,2})
\left[U^\text{st}_{12}(\bq)\right]^2 G_2(P)G_2(P-Q)   \nn \\
	&&\times \left[2G_1(K)G_1(K+Q)-G_1(K)G_1(K+Q+W)-G_1(K+Q)G_1(K+W)\right].
 \label{sigma12_symm_main}
\eea
\ewt
The rest of the calculations is identical to those in Secs.~\ref{sec:intra} and Appendix \ref{sec:comp}, and the final result for the intervalley part reads
\bea
\R\sigma_{\text{inter}}(\omega)&=&-2A_De^2\lambda_{12} \frac{q^{*4}}{\bar m_1\bar m_2 \vfa\vfb \mB^{4-D}}.
\label{sigma12_final}
\eea
Note that $\vfa$ and $\vfb$ in the last equation are the absolute values of the Fermi velocities in the corresponding valleys, while masses $\bar m_1$ and $\bar m_2$ are positive for electron-like valleys and negative for hole-like valleys. Therefore, the intervalley contribution to the conductivity is negative, if both valleys are either electron-like or hole-like, and positive, if one of the valleys is electron-like and and another one is hole-like.
This is exactly the same effect that one encounters in Coulomb drag between two physically separate layers: the sign of the drag conductivity depends on the sign of the relative charge of the carriers in two layers~\cite{Narozhny:2016}. Also note that, like the intravalley contribution, the intervalley one contains only the density-of-states rather than the band mass. This means that the intervalley (drag) contribution is non-zero even for the Dirac single-particle spectrum, for which the band mass vanishes but the density-of-state mass is finite. Therefore, the result is not sensitive to whether the system has a particle-hole symmetry or not.\footnote{The difference between the particle-hole symmetric and asymmetric cases shows up only in subleading terms which contain higher powers of frequency.}

\subsubsection{Total conductivity}
Adding up the intra- and intervalley contributions, Eqs.~\eqref{sigma11_final} and~\eqref{sigma12_final}], we obtain the total conductivity as
\bea
\R\sigma(\omega)&=&\R\sigma_{\text{intra}}(\omega)+\R\sigma_{\text{inter}}(\omega)\nn\\
&=&A_De^2\left(\frac{1}{\bar m_1}-\frac{1}{\bar m_2}\right)^2\lambda_{12} \frac{q^{*4}}{ \vfa\vfb \mB^{4-D}}.\label{sigma_final}
\eea
Note that the conductivity vanishes for the case of identical valleys, i.e., for $\bar m_1=\bar m_2$, in agreement with the result of Sec.~\ref{sec:FGL}.

Employing the scaling argument given at the end of Sec.~\ref{sec:QC2VS}, the frequency dependence of the conductivity is  obtained by replacing $\mB$ by $|\omega|^{1/3}$ in Eq.~\eqref{sigma_final}, which yields
\bea
\R\sigma(\omega)\propto
\left(\frac{1}{\bar m_1}-\frac{1}{\bar m_2}\right)^2\lambda_{12}|\omega|^{-(4-D)/3}.\label{sigma_final_QCP}
\eea
\newline
This recovers the scaling forms, $\omega^{-2/3}$ in 2D~\cite{kim:1994} and $\omega^{-1/3}$ in 3D, obtained under the assumption of a current-relaxing process of an unspecified type.

\section{Conclusions}

We have investigated the impact of Fermi surface (FS) topology and electron-electron interaction (\emph{ee}) interactions on the optical conductivity of a Fermi liquid (FL). Our findings demonstrate that in a FL with a multi-valley FS and distinct effective masses in different valleys, \emph{ee} scattering facilitates a rapid current relaxation due to intervalley drag.

We derived the optical conductivity of a two-valley system, both in the FL regime and near the Ising-nematic quantum-critical point (QCP). Our results show that in the FL regime, intervalley drag restores the Gurzhi scaling form of the conductivity, $\R\sigma(\omega, T) = \sigma_0 \left(1 +  4\pi^2 T^2/\omega^2\right)$. This restoration occurs because the geometric constraints on \emph{ee} scattering are lifted in systems with multiply connected FSs, permitting new scattering channels. Consequently, this effect generates a significant differential response at the threshold of the channel opening, which emerges at the Lifshitz transition from a single-valley to a multi-valley FS\@. We propose seeking this effect in biased bilayer and rhombohedral trilayer graphenes.

Near the QCP, the intervalley contribution to the conductivity scales as $|\omega|^{-2/3}$ in 2D and as $|\omega|^{-1/3}$ in 3D, thus providing a specific current-relaxing process that aligns with longstanding theoretical predictions~\cite{kim:1994}.\\

\begin{acknowledgments}
The work of Y.G.\ was supported by the Simons Foundation Grant No.~1249376. A.L.\ acknowledges financial support by the National Science Foundation (NSF) Grant No. DMR-2203411 and H.I.\ Romnes Faculty Fellowship provided by the University of Wisconsin-Madison Office of the Vice Chancellor for Research and Graduate Education with funding from the Wisconsin Alumni Research Foundation. A.K.\ was supported by the NSF Grant No. DMR-2338819. A.V. Ch.\ was supported the by U.S. Department of Energy, Office of Science, Basic Energy Sciences, under Award No. DE-SC0014402;  D. L. M.  was supported by the NSF grant DMR-2224000.  A. V. Ch., Y.G., A. L., and D. L. M. acknowledge the hospitality of the Kavli Institute for Theoretical Physics, Santa Barbara, supported by the NSF grants PHY-1748958 and PHY-2309135. 
\end{acknowledgments}
 
\onecolumngrid
\appendix
\section{Fermi Golden Rule}
\label{app:FGR}
\begin{figure}[htb]
	\centering
	\begin{subfigure}{.25\linewidth}
	  \centering
	  \includegraphics[width=0.9\linewidth]{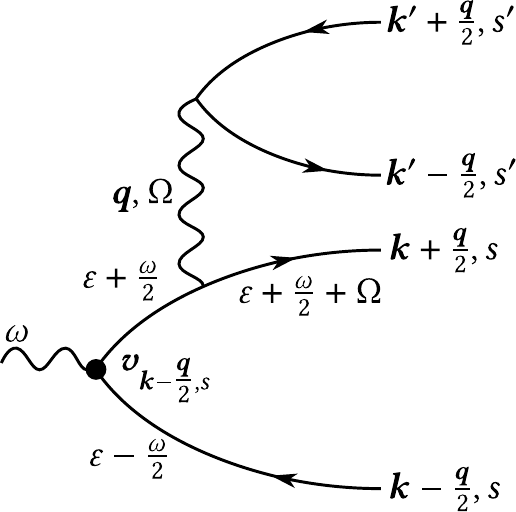}
	  \caption{
	  \label{fig:fgr1}}
	\end{subfigure}%
	\begin{subfigure}{.25\linewidth}
	  \centering
	  \includegraphics[width=0.9\linewidth]{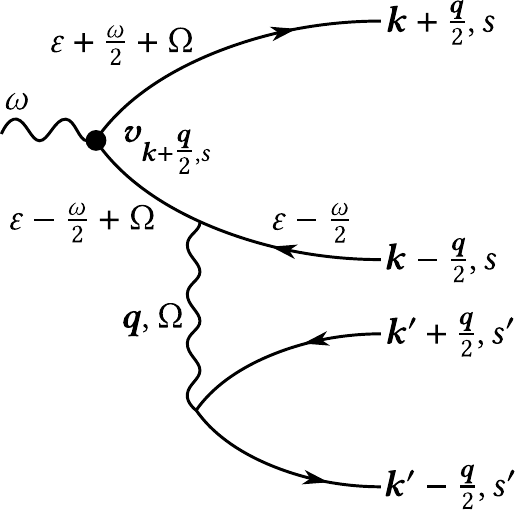}
	 \caption{
	  \label{fig:fgr2}}
	\end{subfigure}%
	\begin{subfigure}{.25\linewidth}
	  \centering
	  \includegraphics[width=0.9\linewidth]{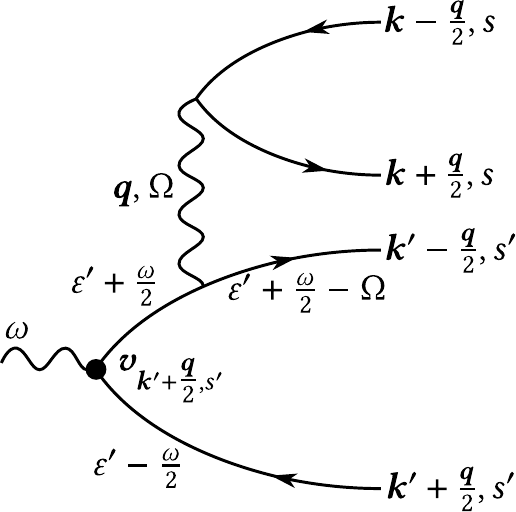}
	  \caption{
	  \label{fig:fgr3}}
	\end{subfigure}%
	\begin{subfigure}{.25\linewidth}
	  \centering
	  \includegraphics[width=0.9\linewidth]{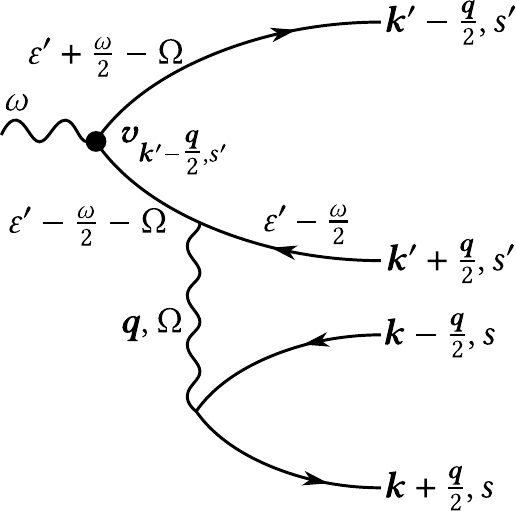}
	 \caption{
	  \label{fig:fgr4}}
	\end{subfigure}
	\RawCaption{\caption{Amplitudes for creation of two electron-hole pairs.}
	\label{fig:fgr}}
\end{figure}
This appendix provides a computation of the imaginary part of the retarded current-current correlator, $\mathcal{K} (\omega)$, to the second order of perturbation theory in \emph{ee} interactions. We use the Källén–Lehmann representation, in essence applying the FGR\@.

An elementary process contributing to $\mathcal{K} (\omega)$ involves the excitation of two electron-hole pairs, one in each of valleys $s$ and $s'$, by a photon of frequency \(\omega\). The contribution is given by:
\begin{equation}
	\delta \mathcal{K}''(\omega) = 2\pi \sum_{\bm{q}} \abs*{\bm{\mathfrak{M}}_{ss'}(\bm{k},\bm{k}',\bm{q})}^2 \delta(\omega +\varepsilon_{\bm{k}-\tfrac{\bm{q}}{2},s} +\varepsilon_{\bm{k}'+\tfrac{\bm{q}}{2},s'}-\varepsilon_{\bm{k}+\tfrac{\bm{q}}{2},s}-\varepsilon_{\bm{k}'-\tfrac{\bm{q}}{2},s'})\,.
\end{equation}
Here, the matrix element $\bm{\mathfrak{M}}_{ss'}(\bm{k},\bm{k}',\bm{q})$ takes into account the virtual states shown in Fig~\ref{fig:fgr}.~\footnote{The exchange processes are disregarded for a long-ranged interaction.} The energy denominator coming from the propagator
$\mathcal{G}_{s} = {(\varepsilon +\dfrac{\omega}{2}-\varepsilon_{\bm{k}-\tfrac{\bm{q}}{2},s})}^{-1}$ simplifies to $\varepsilon +\tfrac{\omega}{2}-(\varepsilon -\tfrac{\omega}{2}) = \omega$ for a process shown in Figs.~\ref{fig:fgr1} (and similarly for \ref{fig:fgr3}), and $-\omega$ for processes in Figs.~\ref{fig:fgr2} and \ref{fig:fgr4}, since the incoming electrons are on the mass shell. Combining all four amplitudes, we get
\begin{equation}
	\bm{\mathfrak{M}}_{ss'}(\bm{k},\bm{k}',\bm{q}) = \frac{u_{ss'}(\bm{q})}{\omega} \left(\bm{v}_{\bm{k}+\tfrac{\bm{q}}{2},s} + \bm{v}_{\bm{k}'-\tfrac{\bm{q}}{2},s'} - \bm{v}_{\bm{k}'+\tfrac{\bm{q}}{2},s'} - \bm{v}_{\bm{k}-\tfrac{\bm{q}}{2},s}\right)\,.
\end{equation}
The total contribution to $\I \mathcal{K} (\omega)$ sums over all electron-hole pair creation and annihilation processes across both valleys, integrating over all initial and final states to give
\begin{align}
    \I \mathcal{K} (\omega) = {} &\frac{2\pi }{4}\sum_{\stackrel{\bm{k} \bm{k'} \bm{q}}{s s'}}
    \abs{\bm{\mathfrak{M}}_{ss'}(\bm{k},\bm{k}',\bm{q})}^2
    \left\{ n_{\bm{k}-\tfrac{\bm{q}}{2},s} n_{\bm{k}'+\tfrac{\bm{q}}{2},s'}
    \left(1-n_{\bm{k}+\tfrac{\bm{q}}{2},s}\right)
    \left(1-n_{\bm{k}'-\tfrac{\bm{q}}{2},s'}\right)
    -\left(1-n_{\bm{k}-\tfrac{\bm{q}}{2},s}\right) \left(1-n_{\bm{k}'+\tfrac{\bm{q}}{2},s'}\right) n_{\bm{k}+\tfrac{\bm{q}}{2},s} n_{\bm{k}'-\tfrac{\bm{q}}{2},s'} \right\} \notag\\
    &{} \times \delta(\omega +\varepsilon_{\bm{k}-\tfrac{\bm{q}}{2},s} +\varepsilon_{\bm{k}'+\tfrac{\bm{q}}{2},s'}-\varepsilon_{\bm{k}+\tfrac{\bm{q}}{2},s}-\varepsilon_{\bm{k}'-\tfrac{\bm{q}}{2},s'})\,,
\end{align}
with equilibrium distribution function $n_{\bm{k}, s}$ accounting for the available states for electron transitions, and $1/4$ accounting for double counting. Finally, the dissipative component of the conductivity, $\R\sigma (\omega)$, is determined via the Kubo formula to give Eq.~\eqref{eq:FGR} of the main text.

\section{Combining the conductivity diagrams}
\label{app:sigma22}

In this Appendix, we demonstrate how the diagrams for the optical conductivity in  Fig.~\ref{fig:sigma11} can be grouped together in such a way that the constraints imposed by momentum conservation become explicit.

\subsection{Self-energy and Maki-Thompson diagrams}
\label{app:abc}

We begin with the self-energy (SE) and Maki-Thompson (MT) diagrams in the intravalley channel,  \emph{a1}-\emph{a3} and  \emph{b1}-\emph{b3} in Fig.~\ref{fig:sigma11}. 
Introducing the fermionic self-energy, $\Sigma_{s}(K)=-\sum_Q G_s(K+Q)U_{ss}(Q)$,  we write the sum of two SE diagrams as
\bea
\sigma_{ss}^{\mathrm{SE}}(\omega_m)=\frac{e^2}{DZ_s^2\omega_m}\sum_K \varv_{s,\bk}^2\left[G^2_{s}(K)\Sigma_s(K) G_s(K+W)+G_{s}(K) G^2_s(K+W)\Sigma_s(K+Q)\right]\Sigma_s(K),\label{SE1}
\eea
where we took into account Eq.~\eqref{Ward} for the current vertex and, as in the main text, $K=(i\nu_m,\bk)$ and $W=(i\omega_m,{\bf 0})$. Using the identity
\bea
G_s(K)G_s(K+W)=\frac{Z_s}{i\omega_m}\left[G_s(K)-G_s(K+Q)\right],
\label{ident}
\eea
Eq.~\eqref{SE1} can be written as
\bea
\sigma_{ss}^{\mathrm{SE}}(\omega_m)=-\frac{1}{D\omega^3_m}\sum_K \varv_{s,\bk}^2\left[G_{s}(K)-G_s(K+W)\right]\left[\Sigma_s(K+W)-\Sigma_s(K)\right].\label{SE2}
\eea
Now we recall that the dissipative part of the conductivity comes only from the dynamic part of the interaction, as defined by Eq.~\eqref{Udyn}. Therefore, $\Sigma_s(K)\to -\sum_K G_s(K+Q) U_{ss}^{\mathrm{dyn}}(Q)$.

Repeating the same steps for the MT diagrams, we obtain for the sum of the SE and MT contributions
\bea
\sigma^{
\text{SE+MT}}_{ss}(\omega_m)&=& - \frac{e^2
}{D\omega^3_m}\sum_{K,Q}
\bv_{\bk,s}\cdot(\bv_{\bk,s}-\bv_{\bk+\bq,s})U_{ss}^{\mathrm{dyn}}(Q)
\nn \\
	&&\times \left[2G_s(K)G_s(K+Q)-G_s(K)G_s(K+Q+W)-G_s(K+Q)G_s(K+W)\right].
 \label{sigma22a-c_nonsymm}
\eea
The expression above can be re-written in a more symmetric form. Relabeling $K+Q\to K$ and then $Q\to -Q$, while keeping in mind that $U_{ss}^{\text{dyn}}(Q)$ is an even function of $Q$, we obtain  an  equivalent form of Eq.~\eqref{sigma22a-c_nonsymm}:
\bea
\sigma^{\text{SE+MT}}_{ss}(\omega_m)
&=& -\frac{e^2
}{D\omega^3_m}\sum_{K,Q}
\bv_{\bk+\bq,s}\cdot(\bv_{\bk+\bq,s}-\bv_{\bk,s})
U_{ss}^{\mathrm{dyn}}(Q)
\nn \\
	&&\times \left[2G_s(K+Q)G_s(K)-G_s(K)G_s(K+Q+W)-G_s(K+Q)G_s(K+W)\right].\label{sigma22a-b_nonsymm_2}
\eea
Taking a half-sum of Eqs.~\eqref{sigma22a-c_nonsymm} and~\eqref{sigma22a-b_nonsymm_2}, we arrive at
\bea
\sigma^{\text{SE+MT}}_{ss}(\omega_m)
&=& -\frac{e^2
}{2D\omega^3_m}\sum_{K,Q}
(\bv_{\bk+\bq,s}-\bv_{\bk,s})^2
U_{ss}^{\mathrm{dyn}}(Q)
\nn \\
	&&\times \left[2G_s(K+Q)G_s(K)-G_s(K+Q)G_s(K+W)-G_s(K+Q)G_s(K+W)\right],\label{sigma22a-c_symm_A}
\eea
which is Eq.~\eqref{sigma22a-c_main} of the main text.
The advantage of the last equation is that the ``transport factor'', $(\bv_{\bk+\bq,s}-\bv_{\bk,s})^2$, which suppresses the contribution from small-$q$ scattering,  is now explicitly quadratic in $q$ for $q\to 0$.

\subsection{Aslamazov-Larkin diagrams}
\label{app:de}
We now turn to the Aslamazov-Larkin (AL) diagrams, considering simultaneously the intravalley  (diagrams \emph{a4}, \emph{a5}, \emph{b4}, and \emph{b5} in Fig.~\ref{fig:sigma11}) and intervalley parts (diagrams \emph{c1-c4} in Fig.~\ref{fig:sigma11}).
\subsubsection{Particle-hole channel}
The particle-hole (\emph{ph}) diagrams \emph{a4}, \emph{b4}, and \emph{c1}
can be written as
\bea
\sigma^{\text{ph}}_{ss'}(\omega_m)&=&-\frac{e^2}{D\omega_m Z_sZ_{s'}} \sum_{K,P,Q}\bv_{\bk,s}\cdot\bv_{\bp-\bq,s'} \left[U^{\text{st}}_{ss'}(\bq)\right]^2
G_s(K)G_s(K+W)G_{s'}(P-Q)G_{s'}(P-Q+W)G_s(K+Q) G_{s'}(P),\label{ALph}
\eea
where $s=s'=1,2$ in diagrams \emph{a4} and  \emph{b4}, and $s=1$, $s'=2$ in diagram \emph{c1}. Diagram \emph{c2} for $\sigma^{\mathrm{ph}}_{21}$ is a mirror image of diagram \emph{c1}, so that $\sigma^{\mathrm{ph}}_{21}(\omega)=\sigma^{\mathrm{ph}}_{12}(\omega)$.

Applying the identity \eqref{ident} to the two first pairs of the Green's functions in Eq.~\eqref{ALph} and opening the brackets, we obtain
\bea
\sigma^{\text{ph}}_{ss'}(\omega_m)&=&\frac{e^2}{D\omega_m^3} \sum_{K,P,Q}\bv_{\bk,s}\cdot\bv_{\bp-\bq,s'} \left[U^{\text{st}}_{ss'}(\bq)\right]^2
\left[G_s(K)-G_s(K+W)\right]\left[G_{s'}(P-Q)-G_{s'}(P-Q+W)\right]G_{s'}(K+Q)G_{s'}(P)\nn\\
&=&\frac{e^2}{D\omega_m^3} \sum_{K,P,Q}\bv_{\bk,s}\cdot\bv_{\bp-\bq,s'} \left[U^{\text{st}}_{ss'}(\bq)\right]^2\left[G_s(K)G_s(K+Q)G_{s'}(P-Q)G_{s'}(P)+G_s(K+W)G_s(K+Q) G_{s'}(P-Q+W)G_{s'}(P)\right.\nn\\
&&\left.- G_s(K)G_s(K+Q) G_{s'}(P-Q+W)G_{s'}(P)-G_s(K+W)G_s(K+Q)  G_{s'}(P-Q)G_{s'}(P)\right].
\eea
In the second term of the last equation we relabel the momenta (under the sum over $K$ and $Q$) as
\bea
&&G_s(K+W)G_s(K+Q) G_{s'}(P-Q+W)G_{s'}(P)\underbrace{\to}_{K+W\leftrightarrow K}G_s(K)G_s(K+Q-W) G_{s'}(P-Q+W)G_{s'}(P)\nn\\
&&\underbrace{\to}_{Q-W\leftrightarrow Q}G_s(K)G_s(K+Q) G_{s'}(P-Q)G_{s'}(P),
\eea
upon which it coincides with the first term. (Note that a replacement $Q-W\leftrightarrow Q$ affects neither the velocity nor static interaction, because $W$ has only the frequency component.)
 Finally, relabeling $Q-W\leftrightarrow Q$ in the third term, we obtain
\bea
\sigma^{\text{ph}}_{ss'}(\omega_m)&=&\frac{e^2}{D\omega_m^3} \sum_{K,P,Q}(\bv_{\bk,s}\cdot\bv_{\bp-\bq,s'})\left[U^{\text{st}}_{ss'}(\bq)\right]^2\left[2G_s(K)G_s(K+Q)-G_s(K)G_s(K+Q+W)-G_s(K+W)G_s(K+Q)\right]\nn\\
&&\times G_{s'}(P-Q)G_{s'}(P).\label{Pid1}
\eea

\subsubsection{Particle-particle channel}
Next, we turn to the particle-particle (\emph{pp}) diagrams \emph{a5}, \emph{b5}, \emph{c3}, and \emph{c4} in Fig.~\ref{fig:sigma11}. Algebraically,
\bea
\sigma^{\text{pp}}_{ss'}(\omega_m)=-\frac{e^2}{D\omega_m Z_sZ_{s'}}\sum_{K,P,Q}(\bv_{s,\bk}\cdot\bv_{s',\bp})\left[U^{\text{st}}_{ss'}(\bq)\right]^2G_s(K) G_s(K+W) G_{s'}(P) G_{s'}(P+W) G_s(K+Q+W) G_{s'}(P-Q),
\eea
where $s=s'=1,2$ for diagrams \emph{a5} and \emph{b5}, and $s=1$, $s'=2$ for diagrams  \emph{c3} and \emph{c4}. As for the particle-hole case, diagram \emph{c4} is a mirror image of \emph{c3}, such that $\sigma^{\text{pp}}_{21}(\omega_m)=\sigma^{\text{pp}}_{12}(\omega_m)$.
Applying Eq.~\eqref{ident} to the first two pairs of Green's functions,   we obtain
\bea
\sigma^{\text{pp}}_{ss'}(\omega_m)&=&\frac{e^2}{D\omega_m^3}\sum_{K,P,Q}(\bv_{s,\bk}\cdot\bv_{s',\bp})\left[U^{\text{st}}_{ss'}(\bq)\right]^2
\left[G_s(K) G_s(K+Q+W) G_{s'}(P) G_{s'}(P-Q)\right.\nn\\
&&\left.+G_s(K+W) G_s(K+Q+W) G_{s'}(P+W) G_{s'}(P-Q)-G_s(K)G_s(K+Q+W) G_{s'}(P+W) G_{s'}(P-Q)\right.\nn\\
&&\left.-G_s(K+W) G_s(K+Q+W) G_{s'}(P) G_{s'}(P-Q)\right].
\eea
In the second and third terms in the last equation, we relabel the momenta (under the sum over $P$ and $Q$) as
\bea
2^{\text{nd}}:\; &&G_s(K+W) G_s(K+Q+W) G_{s'}(P+W) G_{s'}(P-Q)\underbrace{\to}_{P+W\leftrightarrow P} G_s(K+W) G_s(K+Q+W) G_{s'}(P) G_{s'}(P-Q-W)\nn\\
&&\underbrace{\to}_{Q+W\leftrightarrow Q}G_s(K+W) G_s(K+Q) G_{s'}(P) G_{s'}(P-Q),\nn\\
3^{\text{rd}}:\;&&G_s(K)G_s(K+Q+W) G_{s'}(P+W) G_{s'}(P-Q)\underbrace{\to}_{P+W\leftrightarrow P}G_s(K)G_s(K+Q+W) G_{s'}(P)\ G_{s'}(P-Q-W)\nn\\
&&\underbrace{\to}_{Q+W\leftrightarrow Q}G_s(K)G_s(K+Q) G_{s'}(P) G_{s'}(P-Q).
\eea
Finally, relabeling $K+W\leftrightarrow K$ in the fourth term, we obtain the same combination of the Green's functions as in the final expression for the particle-hole channel, Eq.~\eqref{Pid1}, but with an opposite sign:
\bea
\sigma^{\text{pp}}_{ss'}(\omega_m)&=&-\frac{e^2}{D\omega_m^3}\sum_{K,P,Q}(\bv_{s,\bk}\cdot\bv_{s',\bp})\left[2G_s(K) G_s(K+Q)-G_s(K) G_s(K+Q+W)-G_s(K+W) G_s(K+Q)\right]\nn\\
&&\times G_{s'}(P-Q) G_{s'}(P).\label{Pie1}
\eea

\subsubsection{Combined contributions of the particle-hole and particle-particle channels}
The total AL contribution to the conductivity due to intravalley interaction in valleys 1 and 2
is given by the sum of the \emph{ph} and \emph{pp} parts, which are diagonal in valley indices. Adding up equations Eqs.~\eqref{Pid1} and~\eqref{Pie1} with $s=s'$, we find for the intravalley contribution
\bea
\sigma^{\text{AL}}_{ss}(\omega_m)&=&\sigma^{\text{ph}}_{ss}(\omega_m)+\sigma^{\text{pp}}_{ss}(\omega_m)= -\frac{e^2}{D\omega^3_m}\sum_{K,Q}\sum'_P
\bv_{\bk,s}\cdot(\bv_{\bp,s}-\bv_{\bp-\bq,s})
\left[U^\text{st}_{ss}(\bq)\right]^2 G_s(P)G_s(P-Q)   \nn \\
	&&\times \left[2G_s(K)G_s(K+Q)-G_s(K)G_s(K+Q+W)-G_s(K+Q)G_s(K+W)\right],
 \label{sigma22de_nonsymm}
\eea
where the prime over $\sum_P$ indicates that only the dynamic part of the result is to be retained.

The contribution to the conductivity due to intervalley interaction is given by sum of diagrams in panel \emph{c} of Fig.~\ref{fig:sigma11}:
\bea
\sigma_{\text{inter}}(\omega_m)&=&\sigma^{\text{ph}}_{12}(\omega_m)+\sigma^{\text{pp}}_{12}(\omega_m)+\sigma^{\text{ph}}_{21}(\omega_m)+\sigma^{\text{pp}}_{21}(\omega_m)=2\left[\sigma^{\text{ph}}_{12}(\omega_m)+\sigma^{\text{pp}}_{12}(\omega_m)\right]\nn\\
&=&-\frac{2e^2}{D\omega^3_m}\sum_{K,Q}\sum'_P
\bv_{\bk,1}\cdot(\bv_{\bp,2}-\bv_{\bp-\bq,2})
\left[U^\text{st}_{12}(\bq)\right]^2 G_2(P)G_2(P-Q)   \nn \\
	&&\times \left[2G_1(K)G_1(K+Q)-G_1(K)G_1(K+Q+W)-G_1(K+Q)G_1(K+W)\right].
 \label{sigma12_nonsymm}
\eea

\subsubsection{Symmetrized form of the Aslamazov-Larkin contribution}
Like the SE+MT contribution, the AL contribution can also be re-written in a more symmetric form.  To obtain such a form for the intravalley AL contribution, we relabel $K+Q\to K$, $P-Q\to P$, and $Q\to -Q$ in Eq.~\eqref{sigma22de_nonsymm} with the result
\bea
\sigma^{\text{AL}}_{
ss}(\omega_m)&=& -\frac{e^2}{D\omega^3_m}\sum_{K,Q}\sum'_P
\bv_{\bk+\bq,s}\cdot(\bv_{\bp-\bq,s}-\bv_{\bp,s})
\left[U^\text{st}_{ss}(\bq)\right]^2 G_s(P)G_s(P-Q)   \nn \\
	&&\times \left[2G_s(K)G_s(K+Q)-G_s(K)G_s(K+Q+W)-G_s(K+Q)G_s(K+W)\right].
 \label{sigma22de_nonsymm_2}
\eea
Taking a half sum of Eqs.~\eqref{sigma22de_nonsymm} and~\eqref{sigma22de_nonsymm_2}, we arrive at the symmetrized form
\bea
\sigma^{\text{AL}}_{ss}(\omega_m)&=&\frac{e^2}{2D\omega^3_m}\sum_{K,Q}\sum'_P
(\bv_{\bk+\bq,s}-\bv_{\bk,s})\cdot(\bv_{\bp,s}-\bv_{\bp-\bq,s})
\left[U^\text{st}_{ss}(\bq)\right]^2 G_s(P)G_s(P-Q)   \nn \\
	&&\times \left[2G_s(K)G_s(K+Q)-G_s(K)G_s(K+Q+W)-G_s(K+Q)G_s(K+W)\right].
 \label{sigma22de_symm-app}
\eea
As in the SE+MT case, the transport factor in the last equation is manifestly quadratic in $q^2$, which is
 Eq.~\eqref{sigma22de_symm_MT} of the main text.

Likewise, the symmetrized form of the intervalley contribution in Eq.~\eqref{sigma12_nonsymm} can be reduced to
\bea
\sigma_{\text{inter}}(\omega_m)&=&\frac{e^2}{D\omega^3_m}\sum_{K,Q}\sum'_P
(\bv_{\bk+\bq,1}-\bv_{\bk,1})\cdot(\bv_{\bp,2}-\bv_{\bp-\bq,2})
\left[U^\text{st}_{12}(\bq)\right]^2 G_2(P)G_2(P-Q)   \nn \\
	&&\times \left[2G_1(K)G_1(K+Q)-G_1(K)G_1(K+Q+W)-G_1(K+Q)G_1(K+W)\right],
 \label{sigma12_symm}
\eea
which coincides with Eq.~\eqref{sigma12_symm_main} of the main text.
\section{Intravalley contribution to the conductivity of a nearly-critical two-valley system: Computational details}
\label{sec:comp}
In this Appendix, we present a detailed calculation of the conductivity of a nearly-critical two-valley system due to intravalley interaction, given by the sum of Eqs.~\eqref{sigma22a-c_main} and \eqref{sigma22de_symm_MT} of the main text.

At the first step, we integrate over the frequency components of $K$ and $P$, arriving at
\bse
\bea
\sigma^{\text{SE+MT}}_{ss}(\omega_m)&=&-\frac{e^2N_{\mathrm{F},s}^2}{2D\omega^3_m}\int\frac{dqq^{D-1}}{(2\pi)^D}\int\frac{d\Omega_m}{2\pi}\int d\mathcal{O}_{\bq,D}\int\frac{d\mathcal{O}_{\bk,D}}{\mathcal{O}_D}\int d\ve_{\bk,s}
(\bv_{\bk+\bq,s}-\bv_{\bk,s})^2 U_{ss}^\text{dyn}(Q)
\left[n_{\text{F}}(\ve_{\bk,s})-n_{\text{F}}(\ve_{\bk+\bq,s})\right]\nn\\
&&\times\mathcal{G}_s(\bk,\bq,\Omega_m,\Omega_m+\omega_m,\Omega_m-\omega_m),
  \label{sigma11_afternu_SEV} \\
  \sigma^{\text{AL}}_{ss}(\omega_m)&=&
  \frac{e^2 N^2_{\mathrm{F},s}}{2D\omega^3_m}\int\frac{dqq^{D-1}}{(2\pi)^D}\int\frac{d\Omega_m}{2\pi}\int d\mathcal{O}_{\bq,D}\int\frac{d\mathcal{O}_{\bk,D}}{\mathcal{O}_D}\int d\ve_{\bk,s}\int\frac{d\mathcal{O}_{\bp,D}}{\mathcal{O}_D}\int d\ve_{\bp,s}(\bv_{\bk+\bq,s}-\bv_{\bk,s})\cdot(\bv_{\bp,s}-\bv_{\bp-\bq,s})
\left[
U^\text{st}_{ss}(\bq)\right]^2\nn\\
&&\times \left[n_{\text{F}}(\ve_{\bk,s})-n_{\text{F}}(\ve_{\bk+\bq,s})\right]\left[n_{\text{F}}(\ve_{\bp-\bq,s})-n_{\text{F}}(\ve_{\bp,s})\right]\mathcal{G}_s(\bk,\bq,\Omega_m,\Omega_m+\omega_m,\Omega_m-\omega_m) \mathcal{G}_s(\bp-\bq,\bq,\Omega_m,\Omega_m,0),
	\label{sigma11_afternu_AL}
\eea
\ese
where
\bea
\mathcal{G}_s({\bf m},\bq,\omega_1,\omega_2,\omega_3)=Z_s\left(\frac{2}{\frac{i\omega_1}{Z_s}-\ve_{{\bf m}+\bq,s}+\ve_{{\bf m},s}}-\frac{1}{\frac{i\omega_2}{Z_s}-\ve_{{\bf m}+\bq,s}+\ve_{{\bf m},s}}-\frac{1}{\frac{i\omega_3}{Z_s}-\ve_{{\bf m}+\bq,s}+\ve_{{\bf m},s}}\right),\label{calG}
\eea
$N_{\mathrm{F},s}$ is the density of the states of the $s^{\mathrm{th}}$ valley at the corresponding Fermi energy,  $n_F(\ve)$ is the Fermi function, $d\mathcal{O}_{\bk,3}=\sin \phi_\bk d\phi_\bk d\varphi_\bk$ with $\phi_\bk\in(0,\pi)$ and $\varphi_\bk\in(0,2\pi)$ being the polar and azimuthal angles of a spherical system in 3D, respectively, $\mathcal{O}_3=4\pi$, $d\mathcal{O}_{\bk,2}=d\phi_\bk$ with $\phi_\bk\in(-\pi,\pi)$ being the azimuthal angle of a polar system in 2D, and $\mathcal{O}_2=2\pi$.
Now we take into account that typical momenta transfers are small: $q\sim \mB\ll k_{\mathrm{F},s}$. Therefore, the dispersions and the Fermi functions can be expanded to order $\mathcal{O}(q)$:
\bea
\ve_{{\bf k}+\bq,s}-\ve_{{\bf k},s}\approx  \varv_{\text{F},s}q\cos\phi_{\bf k},\,\ve_{{\bf p},s}-\ve_{{\bf p}-\bq,s}\approx  \varv_{\text{F},s}q\cos\phi_{\bf p},\label{ddisp}
\eea
and
\bea
n_{\text{F}}(\ve_{{\bf k},s})-n_{\text{F}}(\ve_{{\bf k}+\bq,s})&\approx& \delta(\ve_{{\bf k},s})\varv_{\text{F},s}
q\cos\phi_{\bf k},\nn\\
n_{\text{F}}(\ve_{{\bf p}-\bq,s})-n_{\text{F}}(\ve_{{\bf p},s})&\approx& \delta(\ve_{{\bf p},s})\varv_{\text{F},s}
q\cos\phi_{\bf p},
\eea
where all the angles are measured from the direction of $\bq$.
Due to the last equation, the integrations over $\ve_{\bk,s}$ and $\ve_{\bp,s}$ project the integrands onto the FS\@.

Next, we turn to the ``transport factors'' in Eqs.~\eqref{sigma11_afternu_SEV} and \eqref{sigma11_afternu_AL}.
For an isotropic but otherwise arbitrary single-particle dispersion, considered in this paper,
 the group velocity in the $s^{\mathrm{th}}$  valley can be written as
\bea
\bv_{\bk,i}=\boldsymbol{\nabla}\ve_{\bk,s}
=
\frac{\bk}{k}\ve_s'(k),
\eea
where $\ve_s(k)\equiv\ve_{\bk,s}$ and $\ve'_s(k)=d\ve_s(k)/dk$.
To linear order in $q$, we have
\bse
\bea
\bv_{\bk+\bq,s}-\bv_{\bk,s}&=& \left(\bq\cdot\boldsymbol{\nabla}\right)\bv_{\bk,s}
= \frac{\bq}{k}\ve'_s(k)+(\hat\bk\cdot\bq)\hat \bk\left(\ve''_s(k)-\frac{\ve'_s(k)}{k}\right),\label{qdelvk}\\
\bv_{\bp,s}-\bv_{\bp-\bq,s}&=& \frac{\bq}{p}\ve'_s(p)+(\hat\bp\cdot\bq)\hat\bp\left(\ve''_s(p)-\frac{\ve'_s(p)}{p}\right),\label{qdelvp}
\eea
\ese
where $\hat\bk=\bk/k$ and $\hat\bp=\bp/p$.
 Therefore, the transport factors, evaluated on the Fermi surface(s), are given by
\bse
\bea
(\bv_{\bk+\bq,s}-\bv_{\bk,s})^2\big\vert_{k=k_{\mathrm{F},s}}&=&q^2\left(\frac{\sin^2\phi_\bk}{\bar m^2_s}+\frac{\cos^2\phi_\bk}{m^{*2}_s}\right),\label{transport}\\
(\bv_{\bk+\bq,s}-\bv_{\bk,s})\cdot(\bv_{\bp,s}-\bv_{\bp-\bq,s})\big\vert_{k=p=k_{\mathrm{F},s}}
&=&q^2\left\{\frac{1}{\bar m_s^2}+(\cos^2\phi_\bk+\cos^2\phi_\bp)\frac{1}{\bar m_s}\left(\frac{1}{m_s^*}-\frac{1}{\bar m_s}\right)
\right.\nn\\
&&\left.+\cos\phi_\bk\cos\phi_\bp\cos\vartheta_{\bk\bp}
\left(\frac{1}{m_s^*}-\frac{1}{\bar m_s}\right)^2\right\},\label{trde_main}
\eea
\ese
where $\vartheta_{\bk\bp}=\angle(\bk,\bp)$, and
\bse
\bea
\frac{1}{\bar m_s}&=&
\frac{1}{k_{\text{F},s}}\ve'_{i}(k)\vert_{k=k_{\text{F},s}}
=\frac{\varv_{\text{F},s}}{k_{\text{F},s}}
\label{mbar}\\
\frac{1}{m^*_s}&=&\ve''_{s}(k)\vert_{k=k_{\text{F},s}}\label{mstar}
\eea
\ese
are the (inverse) density-of-states and band masses of the $s^{\text{th}}$ valley, respectively.

In the FL regime, typical momentum and energy transfers are such that $|\Omega_m|\sim\omega_m\ll \varv_{\text{F},s}q$. Then, as it follows from Eqs.~\eqref{calG} and~\eqref{ddisp}, the relevant values of $\cos\phi_\bk$ and $\cos\phi_\bp$ are small, i.e.,  $|\cos\phi_{\bk,\bp}|\sim \omega_m/\vfa q\ll 1$, which  means that $\phi_{\bk,\bp}\approx \pm \pi/2$ in 2D and $\phi_{\bk,\bp}\approx \pi/2$ in 3D. In this limit, the transport factors in Eqs.~\eqref{transport} and~\eqref{trde_main}
are reduced to to a simpler form
\bea
(\bv_{\bk+\bq,s}-\bv_{\bk,s})^2\big\vert_{k=\kfa}&=&(\bv_{\bk+\bq,s}-\bv_{\bk,s})\cdot(\bv_{\bp,s}-\bv_{\bp-\bq,s})\big\vert_{k=\kfa,p=\kfa}= \frac{q^2}{\bar m_s^2}.
\label{transport_FS}
\eea

The rest of the integrands in Eqs.~\eqref{sigma11_afternu_SEV} and~\eqref{sigma11_afternu_AL},
$\cos\phi_\bk$ and $\cos\phi_\bp$ are linearized near $\phi_{\bk,\bp}=\pm\pi/2$ in 2D and near $\phi_{\bk,\bp}=\pi/2$ in 3D, respectively, and the angular integrals are solved to linear order in $\omega_m/\varv_{\mathrm{F},s} q$. Since $U_{ss}^{\mathrm{dyn}}$ and $\Pi^{\mathrm{dyn}}_{11}$ depend only on the magnitude of $\bq$, the integration over $\mathcal{O}_{\bq,D}$ simply gives a factor of $\mathcal{O}_{D}$. Adding up the SE+MT and AL contributions, we obtain
\bea
\sigma
_{\text{intra}}(\omega_m)&=&\sum_s \frac{e^2C_D \mathcal{O}_{D} N_{\mathrm{F},s}}{2D\omega^3_m
\varv_{\mathrm{F},s}
\bar m_s^2}\int\frac{dqq^D}{(2\pi)^D}
\int\frac{d\Omega_m}{2\pi}
	\left(|\Omega_m+\omega_m|+|\Omega_m-\omega_m|-2|\Omega_m|\right)
\left\{U^{\text{dyn}}_{ss}(\bq,\Omega_m)-\left[U_{ss}^{\text{st}}(\bq)\right]^2\Pi_{ss}^{\text{dyn}}(\bq,\Omega_m)\right\},\nn\\
 \label{sigma22abc_3}
\eea
where $C_2=1$ and $C_3=\pi/2$, which coincides with Eq.~\eqref{sigma22abc_4} of the main text.

\section{``High-energy'' approach to quantum criticality}
\label{sec:high}

In the previous section, we treated all intra- and intervalley interactions on the same footing, within the RPA, while quantum criticality was imposed by tuning the interaction in valley-1 to the critical value. We will refer to this  approach as to a ``low-energy'' one. Another widely accepted way to construct an effective theory of quantum criticality  is  to assume that  critical bosons are formed out of ``high-energy'' fermions with energies comparable to the ultraviolet scale of the model. Bosons are assumed to be described by the Ornstein-Zernike susceptibility
\bea
\chi_0(\bq)=\frac{\mu}{q^2+\mB^2},\label{chi0}
\eea
which incorporates the static part of $\Pi_{11} (\bq)$.  In general, $\chi_0(\bq)$ also contains $\Omega^2$ term coming from the internal dynamics of bosons, but we assume that it is less important than Landau damping by low-energy fermions. We will refer to this approach as to a ``high-energy'' one.

The goal of this section is to show that the high-energy approach has to be treated with extra caution when applied to a two-valley system.  One reason for caution is the renormalization of $\mB$ by intra-valley interaction $u_{12}$. Another, more fundamental reason,  is the need to keep the static $\Pi_{11} (\bq)$ in the formulas. 

Indeed, a formal extension of the high-energy approach to the two-valley case yields the action of the following form:
\bea
S=S_{\text{f}0}+S_{\text{b}0}+S_{\text{fb}}+S_{\text{int},12}+S_{\text{int},22}
\eea
where
\bea
S_{\text{f}0}=\sum_{s=1,2}\int d\tau \sum_\bk \bar c_{\bk,s}(\tau)\left(\partial_\tau+\ve_{\bk,s}\right)c_{\bk,s}(\tau)
\eea
is the action for free fermions  with dispersions $\ve_{\bk,s}
$ ($s=1,2$),
\bea
S_{\text{b}0}=\int d\tau\sum_\bq \frac{\varphi_\bq(\tau)\varphi_{-\bq}(\tau)}{\chi_0(\bq)}\label{Sb0}
\eea
is the the bosonic part,
\bea
S_{\text{fb}}=u\int d\tau\sum_{\bk,\bq}F_{\bk}c^\dagger_{\bk+\frac{\bq}{2},1}(\tau) c_{\bk-\frac{\bq}{2},1}(\tau) \varphi_{-\bq}(\tau)
\eea
describes the coupling of valley-1 fermions to critical bosons, and
\bse
\bea
S_{\text{int},12}&=&u_{12}\int d\tau\sum_{\bk,\bp,\bq}
\bar c_{\bk+\frac{\bq}{2},1}(\tau)\bar c_{\bp-\frac{\bq}{2},2}(\tau)c_{\bp+\frac{\bq}{2},2}(\tau)c_{\bk-\frac{\bq}{2},1}(\tau)
\label{Sint12}
\\
S_{\text{int},22}&=&\frac{u_{22}}{2}\int d\tau
\sum_{\bk,\bp,\bq}
\bar c_{\bk+\frac{\bq}{2},2}(\tau)\bar c_{\bp-\frac{\bq}{2},2}(\tau) c_{\bp+\frac{\bq}{2},2}(\tau)c_{\bk-\frac{\bq}{2},2}(\tau)
\label{Sint22}
\eea
\ese
describe non-critical interactions between valleys 1 and valley 2, and in valley 2, respectively.

Integrating out bosons via the Hubbard-Stratonovich transformation, we obtain
\bea
S_{\text{int},11}&=&\frac 12 \int d\tau\sum_{\bk,\bp,\bq} u_{11}(\bq)
\bar c_{\bk+\frac{\bq}{2},1}(\tau)\bar c_{\bp-\frac{\bq}{2},1}(\tau) c_{\bk-\frac{\bq}{2},1}(\tau) c_{\bp+\frac{\bq}{2},1}(\tau),\label{Sint11}
\eea
where
\bea
u_{11}(\bq)=-u^2\chi_0(\bq)=-\frac{w}{q^2+\mB^2}
\eea
is the $q$-dependent effective interaction with valley 1 with $w=u^2\mu$. 
As we said earlier, the static part of $\Pi_{11} (\bq)$ is incorporated into  $u_{11}(\bq)$ as a portion of $\mB$. 
Other parts of the action are not affected by the Hubbard-Stratonovich transformation. 
The full action can now be written as
\bea
S&=&S_{\text{f}0}+\frac 12\int d\tau\sum_{\bk,\bp,\bq} \sum_{ss'=1,2}u_{ss'}(\bq)\bar c_{\bk+\frac{\bq}{2},s}(\tau)
\bar c_{\bp-\frac{\bq}
{2},s'}(\tau)c_{\bp+\frac{\bq}{2},s'}(\tau)c_{\bk-\frac{\bq}{2},s}(\tau)\label{S}
\eea
where
\bea
\hat u(\bq)=\begin{pmatrix}
u_{11}(\bq) & u_{12} \\
u_{12} & u_{22} \\
\end{pmatrix},
\eea
is a $2\times 2$ matrix of ``bare'' interactions. 
In this formulation, interaction between low-energy fermions introduces Landau damping, but there are no interaction terms that would contain static $\Pi_{11}(\bq)$.  As the consequence, 
if we extract the effective interactions from the 
the action, we would obtain
\bse
\bea
\tilde U_{11}(\bq,\Omega_m)&=&\tilde U_{11,a}(\bq,\Omega_m)+\tilde U_{11,b}(\bq,\Omega_m),\\
\tilde U_{11,a}(\bq,\Omega_m)&=&\frac{
u_{11}(\bq)}{1-u_{11}(\bq)\Pi^{\text{dyn}}_{11}(\bq,\Omega_m)}
=-\frac{w}{q^2+\mB^2+\gamma|\Omega_m|/q},
\label{V11SE}\\
\tilde U_{11,b}(\bq,\Omega_m)&=&\frac{
u_{12}^2\Pi_{22}(\bq,\Omega_m)}{\left[1-u_{11}(\bq)\Pi^\text{dyn}_{11}(\bq,\Omega_m)\right]^2}=-\frac{\Nfb u_{12}^2(q^2+\mB^2)^2}{\left(q^2+\mB^2+\gamma|\Omega_m|/q\right)^2},\\
 \tilde{U}_{12}(\bq,\Omega_m)&=&\frac{u_{12}}{1-u_{11}(\bq)\Pi^{\text{dyn}}_{11}(\bq,\Omega_m)}=u_{12}\frac{q^2+\mB^2}{q^2+\mB^2+\gamma|\Omega_m|/q},\label{V12}\\
\tilde U_{22}(\bq,\Omega_m)&=& \frac{u_{12}^2\Pi^{\text{dyn}}_{11}(\bq,\Omega_m)}{1-u_{11}(\bq)\Pi_{11}^{\text{dyn}}(\bq,\Omega_m)}
=\frac{u_{12}^2}{w}\frac{(q^2+\mB^2)\gamma|\Omega_m|/q
}{q^2+\mB^2+\gamma|\Omega_m|/q}
\label{V22}
\eea
\ese
where $\gamma=C_D w\Nfa/\vfa$ and 
where we neglected the difference between $R_1$ and $R$ in Eqs.~\eqref{R1} and \eqref{R12}.

Comparing with Eqs.~\eqref{U11g}, \eqref{U12g}, and \eqref{U22g} we see that $\tilde U_{11,a}(\bq,\Omega_m)$ is the same as the first term in the last line in Eq.~\eqref{U11g}, but other terms are different. The difference is two-fold. First, all denominators are  powers of  $1-u_{11}(\bq)\Pi^{\text{dyn}}_{11}(\bq,\Omega_m)$, which is the same as $R_1$ in Eq.~\eqref{R1}. The  denominators in Eqs.~\eqref{U11g}, \eqref{U12g}, and \eqref{U22g} contain $R$, which differs from $R_1$ by  $\Pi_{11}(\bq,\Omega_m)\Pi_{22}(\bq,\Omega_m) u_{12}^2$, see Eq.~\eqref{R1}.   Second, the terms in Eq.~\eqref{V22} all contain  small $u_{11}(\bq)$ in the numerator, while the corresponding terms in Eqs.~\eqref{U11g}, \eqref{U12g}, and \eqref{U22g} contain the bare $u_{11}$. The two terms differ by a static $\Pi_{11} (\bq)$: $u^{-1}_{11} (\bq) = u^{-1}_{11} -\Pi_{11} (\bq)$. This static piece is missing if we formally extend the high-energy approach from a one-valley case, where it is valid,  to the two-valley case.

\twocolumngrid
\bibliography{optcondIN}

\begin{thebibliography}{45}%
\makeatletter
\providecommand \@ifxundefined [1]{%
 \@ifx{#1\undefined}
}%
\providecommand \@ifnum [1]{%
 \ifnum #1\expandafter \@firstoftwo
 \else \expandafter \@secondoftwo
 \fi
}%
\providecommand \@ifx [1]{%
 \ifx #1\expandafter \@firstoftwo
 \else \expandafter \@secondoftwo
 \fi
}%
\providecommand \natexlab [1]{#1}%
\providecommand \enquote  [1]{``#1''}%
\providecommand \bibnamefont  [1]{#1}%
\providecommand \bibfnamefont [1]{#1}%
\providecommand \citenamefont [1]{#1}%
\providecommand \href@noop [0]{\@secondoftwo}%
\providecommand \href [0]{\begingroup \@sanitize@url \@href}%
\providecommand \@href[1]{\@@startlink{#1}\@@href}%
\providecommand \@@href[1]{\endgroup#1\@@endlink}%
\providecommand \@sanitize@url [0]{\catcode `\\12\catcode `\$12\catcode `\&12\catcode `\#12\catcode `\^12\catcode `\_12\catcode `\%12\relax}%
\providecommand \@@startlink[1]{}%
\providecommand \@@endlink[0]{}%
\providecommand \url  [0]{\begingroup\@sanitize@url \@url }%
\providecommand \@url [1]{\endgroup\@href {#1}{\urlprefix }}%
\providecommand \urlprefix  [0]{URL }%
\providecommand \Eprint [0]{\href }%
\providecommand \doibase [0]{https://doi.org/}%
\providecommand \selectlanguage [0]{\@gobble}%
\providecommand \bibinfo  [0]{\@secondoftwo}%
\providecommand \bibfield  [0]{\@secondoftwo}%
\providecommand \translation [1]{[#1]}%
\providecommand \BibitemOpen [0]{}%
\providecommand \bibitemStop [0]{}%
\providecommand \bibitemNoStop [0]{.\EOS\space}%
\providecommand \EOS [0]{\spacefactor3000\relax}%
\providecommand \BibitemShut  [1]{\csname bibitem#1\endcsname}%
\let\auto@bib@innerbib\@empty
\bibitem [{\citenamefont {Peres}(2010)}]{Peres:2010}%
  \BibitemOpen
  \bibfield  {author} {\bibinfo {author} {\bibfnamefont {N.~M.~R.}\ \bibnamefont {Peres}},\ }\bibfield  {title} {\bibinfo {title} {Colloquium: The transport properties of graphene: An introduction},\ }\href {https://doi.org/10.1103/RevModPhys.82.2673} {\bibfield  {journal} {\bibinfo  {journal} {Rev. Mod. Phys.}\ }\textbf {\bibinfo {volume} {82}},\ \bibinfo {pages} {2673} (\bibinfo {year} {2010})}\BibitemShut {NoStop}%
\bibitem [{\citenamefont {Gmitra}\ and\ \citenamefont {Fabian}(2015)}]{Gmitra:2015}%
  \BibitemOpen
  \bibfield  {author} {\bibinfo {author} {\bibfnamefont {M.}~\bibnamefont {Gmitra}}\ and\ \bibinfo {author} {\bibfnamefont {J.}~\bibnamefont {Fabian}},\ }\bibfield  {title} {\bibinfo {title} {Graphene on transition-metal dichalcogenides: A platform for proximity spin-orbit physics and optospintronics},\ }\href {https://doi.org/10.1103/PhysRevB.92.155403} {\bibfield  {journal} {\bibinfo  {journal} {Phys. Rev. B}\ }\textbf {\bibinfo {volume} {92}},\ \bibinfo {pages} {155403} (\bibinfo {year} {2015})}\BibitemShut {NoStop}%
\bibitem [{\citenamefont {Manzeli}\ \emph {et~al.}(2017)\citenamefont {Manzeli}, \citenamefont {Ovchinnikov}, \citenamefont {Pasquier}, \citenamefont {Yazyev},\ and\ \citenamefont {Kis}}]{manzeli:2017}%
  \BibitemOpen
  \bibfield  {author} {\bibinfo {author} {\bibfnamefont {S.}~\bibnamefont {Manzeli}}, \bibinfo {author} {\bibfnamefont {D.}~\bibnamefont {Ovchinnikov}}, \bibinfo {author} {\bibfnamefont {D.}~\bibnamefont {Pasquier}}, \bibinfo {author} {\bibfnamefont {O.~V.}\ \bibnamefont {Yazyev}},\ and\ \bibinfo {author} {\bibfnamefont {A.}~\bibnamefont {Kis}},\ }\bibfield  {title} {\bibinfo {title} {{2D} transition metal dichalcogenides},\ }\href {https://doi.org/10.1038/natrevmats.2017.33} {\bibfield  {journal} {\bibinfo  {journal} {Nature Reviews Materials}\ }\textbf {\bibinfo {volume} {2}},\ \bibinfo {pages} {17033} (\bibinfo {year} {2017})}\BibitemShut {NoStop}%
\bibitem [{\citenamefont {Wang}\ \emph {et~al.}(2018)\citenamefont {Wang}, \citenamefont {Chernikov}, \citenamefont {Glazov}, \citenamefont {Heinz}, \citenamefont {Marie}, \citenamefont {Amand},\ and\ \citenamefont {Urbaszek}}]{heinz:2018}%
  \BibitemOpen
  \bibfield  {author} {\bibinfo {author} {\bibfnamefont {G.}~\bibnamefont {Wang}}, \bibinfo {author} {\bibfnamefont {A.}~\bibnamefont {Chernikov}}, \bibinfo {author} {\bibfnamefont {M.~M.}\ \bibnamefont {Glazov}}, \bibinfo {author} {\bibfnamefont {T.~F.}\ \bibnamefont {Heinz}}, \bibinfo {author} {\bibfnamefont {X.}~\bibnamefont {Marie}}, \bibinfo {author} {\bibfnamefont {T.}~\bibnamefont {Amand}},\ and\ \bibinfo {author} {\bibfnamefont {B.}~\bibnamefont {Urbaszek}},\ }\bibfield  {title} {\bibinfo {title} {Colloquium: Excitons in atomically thin transition metal dichalcogenides},\ }\href {https://doi.org/10.1103/RevModPhys.90.021001} {\bibfield  {journal} {\bibinfo  {journal} {Rev. Mod. Phys.}\ }\textbf {\bibinfo {volume} {90}},\ \bibinfo {pages} {021001} (\bibinfo {year} {2018})}\BibitemShut {NoStop}%
\bibitem [{\citenamefont {Hosur}\ and\ \citenamefont {Qi}(2013)}]{Hosur:2013}%
  \BibitemOpen
  \bibfield  {author} {\bibinfo {author} {\bibfnamefont {P.}~\bibnamefont {Hosur}}\ and\ \bibinfo {author} {\bibfnamefont {X.}~\bibnamefont {Qi}},\ }\bibfield  {title} {\bibinfo {title} {{Recent developments in transport phenomena in Weyl semimetals}},\ }\href {https://doi.org/http://dx.doi.org/10.1016/j.crhy.2013.10.010} {\bibfield  {journal} {\bibinfo  {journal} {C. R. Physique}\ }\textbf {\bibinfo {volume} {14}},\ \bibinfo {pages} {857 } (\bibinfo {year} {2013})}\BibitemShut {NoStop}%
\bibitem [{\citenamefont {Armitage}\ \emph {et~al.}(2018)\citenamefont {Armitage}, \citenamefont {Mele},\ and\ \citenamefont {Vishwanath}}]{Armitage:2018b}%
  \BibitemOpen
  \bibfield  {author} {\bibinfo {author} {\bibfnamefont {N.~P.}\ \bibnamefont {Armitage}}, \bibinfo {author} {\bibfnamefont {E.~J.}\ \bibnamefont {Mele}},\ and\ \bibinfo {author} {\bibfnamefont {A.}~\bibnamefont {Vishwanath}},\ }\bibfield  {title} {\bibinfo {title} {{Weyl and Dirac semimetals in three-dimensional solids}},\ }\href {https://doi.org/10.1103/RevModPhys.90.015001} {\bibfield  {journal} {\bibinfo  {journal} {Rev. Mod. Phys.}\ }\textbf {\bibinfo {volume} {90}},\ \bibinfo {pages} {015001} (\bibinfo {year} {2018})}\BibitemShut {NoStop}%
\bibitem [{\citenamefont {Ivchenko}(2005)}]{ivchenko2005optical}%
  \BibitemOpen
  \bibfield  {author} {\bibinfo {author} {\bibfnamefont {E.~L.}\ \bibnamefont {Ivchenko}},\ }\href@noop {} {\emph {\bibinfo {title} {Optical spectroscopy of semiconductor nanostructures}}}\ (\bibinfo  {publisher} {Alpha Science International Ltd.},\ \bibinfo {year} {2005})\BibitemShut {NoStop}%
\bibitem [{\citenamefont {Basov}\ and\ \citenamefont {Timusk}(2005)}]{basov:2005}%
  \BibitemOpen
  \bibfield  {author} {\bibinfo {author} {\bibfnamefont {D.~N.}\ \bibnamefont {Basov}}\ and\ \bibinfo {author} {\bibfnamefont {T.}~\bibnamefont {Timusk}},\ }\bibfield  {title} {\bibinfo {title} {Electrodynamics of high-${T}_{c}$ superconductors},\ }\href {https://doi.org/10.1103/RevModPhys.77.721} {\bibfield  {journal} {\bibinfo  {journal} {Rev. Mod. Phys.}\ }\textbf {\bibinfo {volume} {77}},\ \bibinfo {pages} {721} (\bibinfo {year} {2005})}\BibitemShut {NoStop}%
\bibitem [{\citenamefont {{Basov}}\ \emph {et~al.}(2011)\citenamefont {{Basov}}, \citenamefont {{Averitt}}, \citenamefont {{van der Marel}}, \citenamefont {{Dressel}},\ and\ \citenamefont {{Haule}}}]{basov:2011}%
  \BibitemOpen
  \bibfield  {author} {\bibinfo {author} {\bibfnamefont {D.~N.}\ \bibnamefont {{Basov}}}, \bibinfo {author} {\bibfnamefont {R.~D.}\ \bibnamefont {{Averitt}}}, \bibinfo {author} {\bibfnamefont {D.}~\bibnamefont {{van der Marel}}}, \bibinfo {author} {\bibfnamefont {M.}~\bibnamefont {{Dressel}}},\ and\ \bibinfo {author} {\bibfnamefont {K.}~\bibnamefont {{Haule}}},\ }\bibfield  {title} {\bibinfo {title} {{Electrodynamics of correlated electron materials}},\ }\href {https://doi.org/10.1103/RevModPhys.83.471} {\bibfield  {journal} {\bibinfo  {journal} {Rev. Mod. Phys.}\ }\textbf {\bibinfo {volume} {83}},\ \bibinfo {pages} {471} (\bibinfo {year} {2011})}\BibitemShut {NoStop}%
\bibitem [{\citenamefont {Klingshirn}(2012)}]{klingshirn2012}%
  \BibitemOpen
  \bibfield  {author} {\bibinfo {author} {\bibfnamefont {C.~F.}\ \bibnamefont {Klingshirn}},\ }\href@noop {} {\emph {\bibinfo {title} {Semiconductor optics}}}\ (\bibinfo  {publisher} {Springer Science \& Business Media},\ \bibinfo {year} {2012})\BibitemShut {NoStop}%
\bibitem [{\citenamefont {Maslov}\ and\ \citenamefont {Chubukov}(2017)}]{maslov:2017b}%
  \BibitemOpen
  \bibfield  {author} {\bibinfo {author} {\bibfnamefont {D.~L.}\ \bibnamefont {Maslov}}\ and\ \bibinfo {author} {\bibfnamefont {A.~V.}\ \bibnamefont {Chubukov}},\ }\bibfield  {title} {\bibinfo {title} {Optical response of correlated electron systems},\ }\href {http://iopscience.iop.org/article/10.1088/1361-6633/80/2/026503} {\bibfield  {journal} {\bibinfo  {journal} {Rep. Prog. Phys.}\ }\textbf {\bibinfo {volume} {80}},\ \bibinfo {pages} {026503} (\bibinfo {year} {2017})}\BibitemShut {NoStop}%
\bibitem [{\citenamefont {Armitage}()}]{Armitage:2018}%
  \BibitemOpen
  \bibfield  {author} {\bibinfo {author} {\bibfnamefont {N.~P.}\ \bibnamefont {Armitage}},\ }\href@noop {} {\bibinfo {title} {Electrodynamics of correlated electron systems, lecture notes from the 2008 {Boulder} school for condensed matter and materials physics}},\ \Eprint {https://arxiv.org/abs/0908.1126} {arXiv:0908.1126 [cond-mat.str-el]} \BibitemShut {NoStop}%
\bibitem [{\citenamefont {Tanner}(2019)}]{Tanner:book}%
  \BibitemOpen
  \bibfield  {author} {\bibinfo {author} {\bibfnamefont {D.~B.}\ \bibnamefont {Tanner}},\ }\href@noop {} {\emph {\bibinfo {title} {{Optical Effects in Solids}}}}\ (\bibinfo  {publisher} {Cambridge University Press, Cambridge},\ \bibinfo {year} {2019})\BibitemShut {NoStop}%
\bibitem [{\citenamefont {Gurzhi}(1959)}]{gurzhi:1959}%
  \BibitemOpen
  \bibfield  {author} {\bibinfo {author} {\bibfnamefont {R.~N.}\ \bibnamefont {Gurzhi}},\ }\bibfield  {title} {\bibinfo {title} {{Mutual Electron Correlations in Metal Optics}},\ }\href@noop {} {\bibfield  {journal} {\bibinfo  {journal} {Sov. Phys.--JETP}\ }\textbf {\bibinfo {volume} {35}},\ \bibinfo {pages} {673} (\bibinfo {year} {1959})}\BibitemShut {NoStop}%
\bibitem [{\citenamefont {Rosch}\ and\ \citenamefont {Howell}(2005)}]{rosch:2005}%
  \BibitemOpen
  \bibfield  {author} {\bibinfo {author} {\bibfnamefont {A.}~\bibnamefont {Rosch}}\ and\ \bibinfo {author} {\bibfnamefont {P.~C.}\ \bibnamefont {Howell}},\ }\bibfield  {title} {\bibinfo {title} {Zero-temperature optical conductivity of ultraclean {F}ermi liquids and superconductors},\ }\href {https://doi.org/10.1103/PhysRevB.72.104510} {\bibfield  {journal} {\bibinfo  {journal} {Phys. Rev. B}\ }\textbf {\bibinfo {volume} {72}},\ \bibinfo {pages} {104510} (\bibinfo {year} {2005})}\BibitemShut {NoStop}%
\bibitem [{\citenamefont {Rosch}(2006)}]{rosch:2006}%
  \BibitemOpen
  \bibfield  {author} {\bibinfo {author} {\bibfnamefont {A.}~\bibnamefont {Rosch}},\ }\bibfield  {title} {\bibinfo {title} {Optical conductivity of clean metals},\ }\href {https://doi.org/10.1002/andp.200510203} {\bibfield  {journal} {\bibinfo  {journal} {Annalen der Physik}\ }\textbf {\bibinfo {volume} {15}},\ \bibinfo {pages} {526} (\bibinfo {year} {2006})}\BibitemShut {NoStop}%
\bibitem [{\citenamefont {Sharma}\ \emph {et~al.}(2021)\citenamefont {Sharma}, \citenamefont {Principi},\ and\ \citenamefont {Maslov}}]{Sharma:2021}%
  \BibitemOpen
  \bibfield  {author} {\bibinfo {author} {\bibfnamefont {P.}~\bibnamefont {Sharma}}, \bibinfo {author} {\bibfnamefont {A.}~\bibnamefont {Principi}},\ and\ \bibinfo {author} {\bibfnamefont {D.~L.}\ \bibnamefont {Maslov}},\ }\bibfield  {title} {\bibinfo {title} {{Optical conductivity of a Dirac-Fermi liquid}},\ }\href {https://doi.org/10.1103/PhysRevB.104.045142} {\bibfield  {journal} {\bibinfo  {journal} {Phys. Rev. B}\ }\textbf {\bibinfo {volume} {104}},\ \bibinfo {pages} {045142} (\bibinfo {year} {2021})}\BibitemShut {NoStop}%
\bibitem [{\citenamefont {Guo}\ \emph {et~al.}(2022)\citenamefont {Guo}, \citenamefont {Patel}, \citenamefont {Esterlis},\ and\ \citenamefont {Sachdev}}]{Guo:2022}%
  \BibitemOpen
  \bibfield  {author} {\bibinfo {author} {\bibfnamefont {H.}~\bibnamefont {Guo}}, \bibinfo {author} {\bibfnamefont {A.~A.}\ \bibnamefont {Patel}}, \bibinfo {author} {\bibfnamefont {I.}~\bibnamefont {Esterlis}},\ and\ \bibinfo {author} {\bibfnamefont {S.}~\bibnamefont {Sachdev}},\ }\bibfield  {title} {\bibinfo {title} {Large-${N}$ theory of critical fermi surfaces. {II}. {C}onductivity},\ }\href {https://doi.org/10.1103/PhysRevB.106.115151} {\bibfield  {journal} {\bibinfo  {journal} {Phys. Rev. B}\ }\textbf {\bibinfo {volume} {106}},\ \bibinfo {pages} {115151} (\bibinfo {year} {2022})}\BibitemShut {NoStop}%
\bibitem [{\citenamefont {Li}\ \emph {et~al.}(2023)\citenamefont {Li}, \citenamefont {Sharma}, \citenamefont {Levchenko},\ and\ \citenamefont {Maslov}}]{Li:2023}%
  \BibitemOpen
  \bibfield  {author} {\bibinfo {author} {\bibfnamefont {S.}~\bibnamefont {Li}}, \bibinfo {author} {\bibfnamefont {P.}~\bibnamefont {Sharma}}, \bibinfo {author} {\bibfnamefont {A.}~\bibnamefont {Levchenko}},\ and\ \bibinfo {author} {\bibfnamefont {D.~L.}\ \bibnamefont {Maslov}},\ }\bibfield  {title} {\bibinfo {title} {Optical conductivity of a metal near an {Ising}-nematic quantum critical point},\ }\href {https://doi.org/10.1103/PhysRevB.108.235125} {\bibfield  {journal} {\bibinfo  {journal} {Phys. Rev. B}\ }\textbf {\bibinfo {volume} {108}},\ \bibinfo {pages} {235125} (\bibinfo {year} {2023})}\BibitemShut {NoStop}%
\bibitem [{\citenamefont {Guo}(2023)}]{Guo:2023}%
  \BibitemOpen
  \bibfield  {author} {\bibinfo {author} {\bibfnamefont {H.}~\bibnamefont {Guo}},\ }\href@noop {} {\bibinfo {title} {Fluctuation spectrum of 2+1d critical fermi surface and its application to optical conductivity and hydrodynamics}} (\bibinfo {year} {2023}),\ \Eprint {https://arxiv.org/abs/2311.03458} {arXiv:2311.03458 [cond-mat.str-el]} \BibitemShut {NoStop}%
\bibitem [{\citenamefont {Gindikin}\ and\ \citenamefont {Chubukov}(2024)}]{Gindikin:2024}%
  \BibitemOpen
  \bibfield  {author} {\bibinfo {author} {\bibfnamefont {Y.}~\bibnamefont {Gindikin}}\ and\ \bibinfo {author} {\bibfnamefont {A.~V.}\ \bibnamefont {Chubukov}},\ }\bibfield  {title} {\bibinfo {title} {Fermi surface geometry and optical conductivity of a two-dimensional electron gas near an {Ising}-nematic quantum critical point},\ }\href {https://doi.org/10.1103/PhysRevB.109.115156} {\bibfield  {journal} {\bibinfo  {journal} {Phys. Rev. B}\ }\textbf {\bibinfo {volume} {109}},\ \bibinfo {pages} {115156} (\bibinfo {year} {2024})}\BibitemShut {NoStop}%
\bibitem [{\citenamefont {Gurzhi}\ \emph {et~al.}(1980)\citenamefont {Gurzhi}, \citenamefont {Kopeliovich},\ and\ \citenamefont {Rutkevich}}]{gurzhi:1980}%
  \BibitemOpen
  \bibfield  {author} {\bibinfo {author} {\bibfnamefont {R.~N.}\ \bibnamefont {Gurzhi}}, \bibinfo {author} {\bibfnamefont {A.~I.}\ \bibnamefont {Kopeliovich}},\ and\ \bibinfo {author} {\bibfnamefont {S.~B.}\ \bibnamefont {Rutkevich}},\ }\bibfield  {title} {\bibinfo {title} {Electrical conductivity of two-dimensional metallic systems},\ }\href {http://jetpletters.ru/ps/1427/article_21710.shtml} {\bibfield  {journal} {\bibinfo  {journal} {Sov. Phys.--JETP Lett.}\ }\textbf {\bibinfo {volume} {32}},\ \bibinfo {pages} {336} (\bibinfo {year} {1980})}\BibitemShut {NoStop}%
\bibitem [{\citenamefont {Gurzhi}\ \emph {et~al.}(1982)\citenamefont {Gurzhi}, \citenamefont {Kopeliovich},\ and\ \citenamefont {Rutkevich}}]{gurzhi:1982}%
  \BibitemOpen
  \bibfield  {author} {\bibinfo {author} {\bibfnamefont {R.}~\bibnamefont {Gurzhi}}, \bibinfo {author} {\bibfnamefont {A.}~\bibnamefont {Kopeliovich}},\ and\ \bibinfo {author} {\bibfnamefont {S.~B.}\ \bibnamefont {Rutkevich}},\ }\bibfield  {title} {\bibinfo {title} {Electric conductivity of two-dimensional metallic systems},\ }\href {http://jetp.ras.ru/cgi-bin/e/index/e/56/1/p159?a=list} {\bibfield  {journal} {\bibinfo  {journal} {Sov. Phys.--JETP}\ }\textbf {\bibinfo {volume} {56}},\ \bibinfo {pages} {159} (\bibinfo {year} {1982})}\BibitemShut {NoStop}%
\bibitem [{\citenamefont {{Gurzhi}}\ \emph {et~al.}(1987)\citenamefont {{Gurzhi}}, \citenamefont {{Kopeliovich}},\ and\ \citenamefont {{Rutkevich}}}]{gurzhi:1987}%
  \BibitemOpen
  \bibfield  {author} {\bibinfo {author} {\bibfnamefont {R.~N.}\ \bibnamefont {{Gurzhi}}}, \bibinfo {author} {\bibfnamefont {A.~I.}\ \bibnamefont {{Kopeliovich}}},\ and\ \bibinfo {author} {\bibfnamefont {S.~B.}\ \bibnamefont {{Rutkevich}}},\ }\bibfield  {title} {\bibinfo {title} {{Kinetic properties of two-dimensional metal systems}},\ }\href {https://doi.org/10.1080/00018738700101002} {\bibfield  {journal} {\bibinfo  {journal} {Adv. Phys.}\ }\textbf {\bibinfo {volume} {36}},\ \bibinfo {pages} {221} (\bibinfo {year} {1987})}\BibitemShut {NoStop}%
\bibitem [{\citenamefont {Gurzhi}\ \emph {et~al.}(1995)\citenamefont {Gurzhi}, \citenamefont {Kalinenko},\ and\ \citenamefont {Kopeliovich}}]{gurzhi:1995}%
  \BibitemOpen
  \bibfield  {author} {\bibinfo {author} {\bibfnamefont {R.~N.}\ \bibnamefont {Gurzhi}}, \bibinfo {author} {\bibfnamefont {A.~N.}\ \bibnamefont {Kalinenko}},\ and\ \bibinfo {author} {\bibfnamefont {A.~I.}\ \bibnamefont {Kopeliovich}},\ }\bibfield  {title} {\bibinfo {title} {Electron-electron momentum relaxation in a two-dimensional electron gas},\ }\href {https://doi.org/10.1103/PhysRevB.52.4744} {\bibfield  {journal} {\bibinfo  {journal} {Phys. Rev. B}\ }\textbf {\bibinfo {volume} {52}},\ \bibinfo {pages} {4744} (\bibinfo {year} {1995})}\BibitemShut {NoStop}%
\bibitem [{\citenamefont {Ledwith}\ \emph {et~al.}(2019)\citenamefont {Ledwith}, \citenamefont {Guo},\ and\ \citenamefont {Levitov}}]{levitov:2019}%
  \BibitemOpen
  \bibfield  {author} {\bibinfo {author} {\bibfnamefont {P.~J.}\ \bibnamefont {Ledwith}}, \bibinfo {author} {\bibfnamefont {H.}~\bibnamefont {Guo}},\ and\ \bibinfo {author} {\bibfnamefont {L.}~\bibnamefont {Levitov}},\ }\bibfield  {title} {\bibinfo {title} {The hierarchy of excitation lifetimes in two-dimensional {F}ermi gases},\ }\href {https://doi.org/https://doi.org/10.1016/j.aop.2019.167913} {\bibfield  {journal} {\bibinfo  {journal} {Ann. Phys.}\ }\textbf {\bibinfo {volume} {411}},\ \bibinfo {pages} {167913} (\bibinfo {year} {2019})}\BibitemShut {NoStop}%
\bibitem [{\citenamefont {Mishchenko}\ \emph {et~al.}(2004)\citenamefont {Mishchenko}, \citenamefont {Reizer},\ and\ \citenamefont {Glazman}}]{mishchenko:2004}%
  \BibitemOpen
  \bibfield  {author} {\bibinfo {author} {\bibfnamefont {E.~G.}\ \bibnamefont {Mishchenko}}, \bibinfo {author} {\bibfnamefont {M.~Y.}\ \bibnamefont {Reizer}},\ and\ \bibinfo {author} {\bibfnamefont {L.~I.}\ \bibnamefont {Glazman}},\ }\bibfield  {title} {\bibinfo {title} {Plasmon attenuation and optical conductivity of a two-dimensional electron gas},\ }\href {https://doi.org/10.1103/PhysRevB.69.195302} {\bibfield  {journal} {\bibinfo  {journal} {Phys. Rev. B}\ }\textbf {\bibinfo {volume} {69}},\ \bibinfo {pages} {195302} (\bibinfo {year} {2004})}\BibitemShut {NoStop}%
\bibitem [{\citenamefont {Sharma}\ \emph {et~al.}(2024)\citenamefont {Sharma}, \citenamefont {Principi}, \citenamefont {Vignale},\ and\ \citenamefont {Maslov}}]{Sharma:2024}%
  \BibitemOpen
  \bibfield  {author} {\bibinfo {author} {\bibfnamefont {P.}~\bibnamefont {Sharma}}, \bibinfo {author} {\bibfnamefont {A.}~\bibnamefont {Principi}}, \bibinfo {author} {\bibfnamefont {G.}~\bibnamefont {Vignale}},\ and\ \bibinfo {author} {\bibfnamefont {D.~L.}\ \bibnamefont {Maslov}},\ }\bibfield  {title} {\bibinfo {title} {Optical conductivity and damping of plasmons due to electron-electron interaction},\ }\href {https://doi.org/10.1103/PhysRevB.109.045431} {\bibfield  {journal} {\bibinfo  {journal} {Phys. Rev. B}\ }\textbf {\bibinfo {volume} {109}},\ \bibinfo {pages} {045431} (\bibinfo {year} {2024})}\BibitemShut {NoStop}%
\bibitem [{\citenamefont {Maslov}\ \emph {et~al.}(2011)\citenamefont {Maslov}, \citenamefont {Yudson},\ and\ \citenamefont {Chubukov}}]{maslov:2011}%
  \BibitemOpen
  \bibfield  {author} {\bibinfo {author} {\bibfnamefont {D.~L.}\ \bibnamefont {Maslov}}, \bibinfo {author} {\bibfnamefont {V.~I.}\ \bibnamefont {Yudson}},\ and\ \bibinfo {author} {\bibfnamefont {A.~V.}\ \bibnamefont {Chubukov}},\ }\bibfield  {title} {\bibinfo {title} {{Resistivity of a Non-Galilean--Invariant Fermi Liquid near Pomeranchuk Quantum Criticality}},\ }\href {https://doi.org/10.1103/PhysRevLett.106.106403} {\bibfield  {journal} {\bibinfo  {journal} {Phys. Rev. Lett.}\ }\textbf {\bibinfo {volume} {106}},\ \bibinfo {pages} {106403} (\bibinfo {year} {2011})}\BibitemShut {NoStop}%
\bibitem [{\citenamefont {Pal}\ \emph {et~al.}(2012)\citenamefont {Pal}, \citenamefont {Yudson},\ and\ \citenamefont {Maslov}}]{pal:2012b}%
  \BibitemOpen
  \bibfield  {author} {\bibinfo {author} {\bibfnamefont {H.~K.}\ \bibnamefont {Pal}}, \bibinfo {author} {\bibfnamefont {V.~I.}\ \bibnamefont {Yudson}},\ and\ \bibinfo {author} {\bibfnamefont {D.~L.}\ \bibnamefont {Maslov}},\ }\bibfield  {title} {\bibinfo {title} {{Resistivity of non-Galilean-invariant Fermi- and non-Fermi liquids}},\ }\href@noop {} {\bibfield  {journal} {\bibinfo  {journal} {Lith. J. Phys.}\ }\textbf {\bibinfo {volume} {52}},\ \bibinfo {pages} {142} (\bibinfo {year} {2012})}\BibitemShut {NoStop}%
\bibitem [{\citenamefont {Gantmakher}\ and\ \citenamefont {Levinson}(1987)}]{levinson:book}%
  \BibitemOpen
  \bibfield  {author} {\bibinfo {author} {\bibfnamefont {V.~F.}\ \bibnamefont {Gantmakher}}\ and\ \bibinfo {author} {\bibfnamefont {Y.~B.}\ \bibnamefont {Levinson}},\ }\href@noop {} {\emph {\bibinfo {title} {Carrier Scattering in Metals and Semiconductors}}}\ (\bibinfo  {publisher} {North-Holland, Amsterdam},\ \bibinfo {year} {1987})\BibitemShut {NoStop}%
\bibitem [{\citenamefont {Fradkin}\ \emph {et~al.}(2010)\citenamefont {Fradkin}, \citenamefont {Kivelson}, \citenamefont {Lawler}, \citenamefont {Eisenstein},\ and\ \citenamefont {Mackenzie}}]{fradkin:2010}%
  \BibitemOpen
  \bibfield  {author} {\bibinfo {author} {\bibfnamefont {E.}~\bibnamefont {Fradkin}}, \bibinfo {author} {\bibfnamefont {S.~A.}\ \bibnamefont {Kivelson}}, \bibinfo {author} {\bibfnamefont {M.~J.}\ \bibnamefont {Lawler}}, \bibinfo {author} {\bibfnamefont {J.~P.}\ \bibnamefont {Eisenstein}},\ and\ \bibinfo {author} {\bibfnamefont {A.~P.}\ \bibnamefont {Mackenzie}},\ }\bibfield  {title} {\bibinfo {title} {{Nematic Fermi Fluids in Condensed Matter Physics}},\ }\href {https://doi.org/10.1146/annurev-conmatphys-070909-103925} {\bibfield  {journal} {\bibinfo  {journal} {Ann. Rev. Cond. Mat. Phys.}\ }\textbf {\bibinfo {volume} {{\bf 1}}},\ \bibinfo {pages} {153} (\bibinfo {year} {2010})}\BibitemShut {NoStop}%
\bibitem [{\citenamefont {Kim}\ \emph {et~al.}(1994)\citenamefont {Kim}, \citenamefont {Furusaki}, \citenamefont {Wen},\ and\ \citenamefont {Lee}}]{kim:1994}%
  \BibitemOpen
  \bibfield  {author} {\bibinfo {author} {\bibfnamefont {Y.~B.}\ \bibnamefont {Kim}}, \bibinfo {author} {\bibfnamefont {A.}~\bibnamefont {Furusaki}}, \bibinfo {author} {\bibfnamefont {X.-G.}\ \bibnamefont {Wen}},\ and\ \bibinfo {author} {\bibfnamefont {P.~A.}\ \bibnamefont {Lee}},\ }\bibfield  {title} {\bibinfo {title} {Gauge-invariant response functions of fermions coupled to a gauge field},\ }\href {https://doi.org/10.1103/PhysRevB.50.17917} {\bibfield  {journal} {\bibinfo  {journal} {Phys. Rev. B}\ }\textbf {\bibinfo {volume} {50}},\ \bibinfo {pages} {17917} (\bibinfo {year} {1994})}\BibitemShut {NoStop}%
\bibitem [{\citenamefont {Bir}\ and\ \citenamefont {Pikus}(1974)}]{Bir1974}%
  \BibitemOpen
  \bibfield  {author} {\bibinfo {author} {\bibfnamefont {G.~L.}\ \bibnamefont {Bir}}\ and\ \bibinfo {author} {\bibfnamefont {G.~E.}\ \bibnamefont {Pikus}},\ }\href {https://books.google.com/books/about/Symmetry_and_Strain_induced_Effects_in_S.html?id=38m2QgAACAAJ} {\emph {\bibinfo {title} {{Symmetry and Strain-induced Effects in Semiconductors}}}}\ (\bibinfo  {publisher} {Wiley},\ \bibinfo {address} {Hoboken, NJ, USA},\ \bibinfo {year} {1974})\BibitemShut {NoStop}%
\bibitem [{\citenamefont {Goyal}\ \emph {et~al.}(2023)\citenamefont {Goyal}, \citenamefont {Sharma},\ and\ \citenamefont {Maslov}}]{Goyal:2023}%
  \BibitemOpen
  \bibfield  {author} {\bibinfo {author} {\bibfnamefont {A.~P.}\ \bibnamefont {Goyal}}, \bibinfo {author} {\bibfnamefont {P.}~\bibnamefont {Sharma}},\ and\ \bibinfo {author} {\bibfnamefont {D.~L.}\ \bibnamefont {Maslov}},\ }\bibfield  {title} {\bibinfo {title} {{Intrinsic optical absorption in Dirac metals}},\ }\href {https://doi.org/https://doi.org/10.1016/j.aop.2023.169355} {\bibfield  {journal} {\bibinfo  {journal} {Ann. Phys.}\ ,\ \bibinfo {pages} {169355}} (\bibinfo {year} {2023})}\BibitemShut {NoStop}%
\bibitem [{\citenamefont {McCann}\ and\ \citenamefont {Koshino}(2013)}]{McCann_2013}%
  \BibitemOpen
  \bibfield  {author} {\bibinfo {author} {\bibfnamefont {E.}~\bibnamefont {McCann}}\ and\ \bibinfo {author} {\bibfnamefont {M.}~\bibnamefont {Koshino}},\ }\bibfield  {title} {\bibinfo {title} {The electronic properties of bilayer graphene},\ }\href {https://doi.org/10.1088/0034-4885/76/5/056503} {\bibfield  {journal} {\bibinfo  {journal} {Rep. Prog. Phys.}\ }\textbf {\bibinfo {volume} {76}},\ \bibinfo {pages} {056503} (\bibinfo {year} {2013})}\BibitemShut {NoStop}%
\bibitem [{\citenamefont {Ghazaryan}\ \emph {et~al.}(2023)\citenamefont {Ghazaryan}, \citenamefont {Holder}, \citenamefont {Berg},\ and\ \citenamefont {Serbyn}}]{PhysRevB.107.104502}%
  \BibitemOpen
  \bibfield  {author} {\bibinfo {author} {\bibfnamefont {A.}~\bibnamefont {Ghazaryan}}, \bibinfo {author} {\bibfnamefont {T.}~\bibnamefont {Holder}}, \bibinfo {author} {\bibfnamefont {E.}~\bibnamefont {Berg}},\ and\ \bibinfo {author} {\bibfnamefont {M.}~\bibnamefont {Serbyn}},\ }\bibfield  {title} {\bibinfo {title} {Multilayer graphenes as a platform for interaction-driven physics and topological superconductivity},\ }\href {https://doi.org/10.1103/PhysRevB.107.104502} {\bibfield  {journal} {\bibinfo  {journal} {Phys. Rev. B}\ }\textbf {\bibinfo {volume} {107}},\ \bibinfo {pages} {104502} (\bibinfo {year} {2023})}\BibitemShut {NoStop}%
\bibitem [{\citenamefont {Voigtl{\"a}nder}(2015)}]{voigtlander2015}%
  \BibitemOpen
  \bibfield  {author} {\bibinfo {author} {\bibfnamefont {B.}~\bibnamefont {Voigtl{\"a}nder}},\ }\href@noop {} {\emph {\bibinfo {title} {Scanning probe microscopy: Atomic force microscopy and scanning tunneling microscopy}}}\ (\bibinfo  {publisher} {Springer},\ \bibinfo {year} {2015})\BibitemShut {NoStop}%
\bibitem [{\citenamefont {Maslov}\ \emph {et~al.}(2017)\citenamefont {Maslov}, \citenamefont {Sharma}, \citenamefont {Torbunov},\ and\ \citenamefont {Chubukov}}]{maslov:2017}%
  \BibitemOpen
  \bibfield  {author} {\bibinfo {author} {\bibfnamefont {D.~L.}\ \bibnamefont {Maslov}}, \bibinfo {author} {\bibfnamefont {P.}~\bibnamefont {Sharma}}, \bibinfo {author} {\bibfnamefont {D.}~\bibnamefont {Torbunov}},\ and\ \bibinfo {author} {\bibfnamefont {A.~V.}\ \bibnamefont {Chubukov}},\ }\bibfield  {title} {\bibinfo {title} {Gradient terms in quantum-critical theories of itinerant fermions},\ }\href {https://doi.org/10.1103/PhysRevB.96.085137} {\bibfield  {journal} {\bibinfo  {journal} {Phys. Rev. B}\ }\textbf {\bibinfo {volume} {96}},\ \bibinfo {pages} {085137} (\bibinfo {year} {2017})}\BibitemShut {NoStop}%
\bibitem [{\citenamefont {Hertz}(1976)}]{hertz:1976}%
  \BibitemOpen
  \bibfield  {author} {\bibinfo {author} {\bibfnamefont {J.~A.}\ \bibnamefont {Hertz}},\ }\bibfield  {title} {\bibinfo {title} {Quantum critical phenomena},\ }\href {https://doi.org/10.1103/PhysRevB.14.1165} {\bibfield  {journal} {\bibinfo  {journal} {Phys. Rev. B}\ }\textbf {\bibinfo {volume} {14}},\ \bibinfo {pages} {1165} (\bibinfo {year} {1976})}\BibitemShut {NoStop}%
\bibitem [{\citenamefont {Pimenov}\ \emph {et~al.}(2022)\citenamefont {Pimenov}, \citenamefont {Kamenev},\ and\ \citenamefont {Chubukov}}]{Pimenov:2022}%
  \BibitemOpen
  \bibfield  {author} {\bibinfo {author} {\bibfnamefont {D.}~\bibnamefont {Pimenov}}, \bibinfo {author} {\bibfnamefont {A.}~\bibnamefont {Kamenev}},\ and\ \bibinfo {author} {\bibfnamefont {A.~V.}\ \bibnamefont {Chubukov}},\ }\bibfield  {title} {\bibinfo {title} {Quasiparticle scattering in a superconductor near a nematic critical point: Resonance mode and multiple attractive channels},\ }\href {https://doi.org/10.1103/PhysRevLett.128.017001} {\bibfield  {journal} {\bibinfo  {journal} {Phys. Rev. Lett.}\ }\textbf {\bibinfo {volume} {128}},\ \bibinfo {pages} {017001} (\bibinfo {year} {2022})}\BibitemShut {NoStop}%
\bibitem [{\citenamefont {Chubukov}\ and\ \citenamefont {Maslov}(2017)}]{chubukov:2017}%
  \BibitemOpen
  \bibfield  {author} {\bibinfo {author} {\bibfnamefont {A.~V.}\ \bibnamefont {Chubukov}}\ and\ \bibinfo {author} {\bibfnamefont {D.~L.}\ \bibnamefont {Maslov}},\ }\bibfield  {title} {\bibinfo {title} {Optical conductivity of a two-dimensional metal near a quantum critical point: The status of the extended {D}rude formula},\ }\href {https://doi.org/10.1103/PhysRevB.96.205136} {\bibfield  {journal} {\bibinfo  {journal} {Phys. Rev. B}\ }\textbf {\bibinfo {volume} {96}},\ \bibinfo {pages} {205136} (\bibinfo {year} {2017})}\BibitemShut {NoStop}%
\bibitem [{\citenamefont {Eberlein}\ \emph {et~al.}(2016)\citenamefont {Eberlein}, \citenamefont {Mandal},\ and\ \citenamefont {Sachdev}}]{eberlein:2016}%
  \BibitemOpen
  \bibfield  {author} {\bibinfo {author} {\bibfnamefont {A.}~\bibnamefont {Eberlein}}, \bibinfo {author} {\bibfnamefont {I.}~\bibnamefont {Mandal}},\ and\ \bibinfo {author} {\bibfnamefont {S.}~\bibnamefont {Sachdev}},\ }\bibfield  {title} {\bibinfo {title} {Hyperscaling violation at the ising-nematic quantum critical point in two-dimensional metals},\ }\href {https://doi.org/10.1103/PhysRevB.94.045133} {\bibfield  {journal} {\bibinfo  {journal} {Phys. Rev. B}\ }\textbf {\bibinfo {volume} {94}},\ \bibinfo {pages} {045133} (\bibinfo {year} {2016})}\BibitemShut {NoStop}%
\bibitem [{\citenamefont {{Chubukov}}(2005)}]{Chubukov:2005self}%
  \BibitemOpen
  \bibfield  {author} {\bibinfo {author} {\bibfnamefont {A.~V.}\ \bibnamefont {{Chubukov}}},\ }\bibfield  {title} {\bibinfo {title} {{Self-generated locality near a ferromagnetic quantum critical point}},\ }\href {https://doi.org/10.1103/PhysRevB.71.245123} {\bibfield  {journal} {\bibinfo  {journal} {\prb}\ }\textbf {\bibinfo {volume} {71}},\ \bibinfo {eid} {245123} (\bibinfo {year} {2005})}\BibitemShut {NoStop}%
\bibitem [{\citenamefont {Narozhny}\ and\ \citenamefont {Levchenko}(2016)}]{Narozhny:2016}%
  \BibitemOpen
  \bibfield  {author} {\bibinfo {author} {\bibfnamefont {B.~N.}\ \bibnamefont {Narozhny}}\ and\ \bibinfo {author} {\bibfnamefont {A.}~\bibnamefont {Levchenko}},\ }\bibfield  {title} {\bibinfo {title} {Coulomb drag},\ }\href {https://doi.org/10.1103/RevModPhys.88.025003} {\bibfield  {journal} {\bibinfo  {journal} {Rev. Mod. Phys.}\ }\textbf {\bibinfo {volume} {88}},\ \bibinfo {pages} {025003} (\bibinfo {year} {2016})}\BibitemShut {NoStop}%
\end{thebibliography}%
	
\end{document}